\documentclass[iop]{emulateapj}
\slugcomment{Accepted in AJ - 16 Feb. 2015}
\usepackage{hyperref}
\usepackage{breakurl}
\usepackage{color}
\shorttitle{APOGEE Spectroscopy of the \emph{Kepler} Field}
\shortauthors{Scott W. Fleming et al.}
\begin{document}
\title{The APOGEE Spectroscopic Survey of \emph{Kepler} Planet Hosts: Feasibility, Efficiency, and First Results}

\author{Scott W. Fleming\altaffilmark{1,2},
Suvrath Mahadevan\altaffilmark{3,4},
Rohit Deshpande\altaffilmark{3,4},
Chad F. Bender\altaffilmark{3,4},
Ryan C. Terrien\altaffilmark{3,4},
Robert C. Marchwinski\altaffilmark{3,4},
Ji Wang\altaffilmark{5},
Arpita Roy\altaffilmark{3,4},
Keivan G. Stassun\altaffilmark{6,7},
Carlos Allende Prieto\altaffilmark{8,9},
Katia Cunha\altaffilmark{10,11},
Verne V. Smith\altaffilmark{12},
Eric Agol\altaffilmark{13},
Hasan Ak\altaffilmark{14,3},
Fabienne A. Bastien\altaffilmark{3,15},
Dmitry Bizyaev\altaffilmark{16},
Justin R. Crepp\altaffilmark{17},
Eric B. Ford\altaffilmark{3,4},
Peter M. Frinchaboy\altaffilmark{18},
Domingo An{\'{\i}}bal Garc{\'{\i}}a-Hern{\'{a}}ndez\altaffilmark{8,9},
Ana Elia Garc{\'{\i}}a P{\'e}rez\altaffilmark{19},
B. Scott Gaudi\altaffilmark{20},
Jian Ge\altaffilmark{21},
Fred Hearty\altaffilmark{3},
Bo Ma\altaffilmark{21},
Steve R. Majewski\altaffilmark{19},
Szabolcs M{\'e}sz{\'a}ros\altaffilmark{22},
David L. Nidever\altaffilmark{23,19},
Kaike Pan\altaffilmark{16},
Joshua Pepper\altaffilmark{24,6},
Marc H. Pinsonneault\altaffilmark{20},
Ricardo P. Schiavon\altaffilmark{25},
Donald P. Schneider\altaffilmark{3,4},
John C. Wilson\altaffilmark{19},
Olga Zamora\altaffilmark{8,9},
Gail Zasowski\altaffilmark{26,27}}

\email{fleming@stsci.edu}
\altaffiltext{1}{Computer Science Corporation, 3700 San Martin Dr, Baltimore, MD, 21218, USA}
\altaffiltext{2}{Space Telescope Science Institute, 3700 San Martin Dr, Baltimore, MD, 21218, USA}
\altaffiltext{3}{Department of Astronomy and Astrophysics, The Pennsylvania State University, 525 Davey Laboratory, University Park, PA 16802, USA}
\altaffiltext{4}{Center for Exoplanets and Habitable Worlds, The Pennsylvania State University, University Park, PA 16802, USA}
\altaffiltext{5}{Department of Astronomy, Yale University, New Haven, CT 06511, USA}
\altaffiltext{6}{Department of Physics \& Astronomy, Vanderbilt University, Nashville, TN 37235, USA}
\altaffiltext{7}{Department of Physics, Fisk University, Nashville, TN 37208, USA}
\altaffiltext{8}{Instituto de Astrof{\'{\i}}sica de Canarias (IAC), E-38200 La Laguna, Tenerife, Spain}
\altaffiltext{9}{Departamento de Astrof{\'{\i}}sica, Universidad de La Laguna, E-38206 la Laguna, Tenerife, Spain}
\altaffiltext{10}{Observatorio Nacional-MCTI, Rio de Janeiro, RJ 20921-400, Brazil}
\altaffiltext{11}{Steward Observatory, University of Arizona, Tucson, AZ 85721, USA}
\altaffiltext{12}{National Optical Astronomy Observatory, 950 N. Cherry Avenue, Tucson, AZ, 85719, USA}
\altaffiltext{13}{Department of Astronomy, Box 351580, University of Washington, Seattle, WA 98195, USA}
\altaffiltext{14}{Faculty of Sciences, Department of Astronomy and Space Sciences, Erciyes University, 38039 Kayseri, Turkey}
\altaffiltext{15}{Hubble Fellow}
\altaffiltext{16}{Apache Point Observatory, P.O. Box 59, Sunspot, NM 88349-0059, USA}
\altaffiltext{17}{University of Notre Dame, Department of Physics, 225 Nieuwland Science Hall, Notre Dame, IN 46556, USA}
\altaffiltext{18}{Department of Physics and Astronomy, Texas Christian University, Fort Worth, TX 76129, USA}
\altaffiltext{19}{Department of Astronomy, University of Virginia, Charlottesville, VA 22904-4325, USA}
\altaffiltext{20}{Department of Astronomy, The Ohio State University, 140 West 18th Avenue, Columbus, OH 43210, USA}
\altaffiltext{21}{Astronomy Department, University of Florida, 211 Bryant Space Sciences Center, Gainesville, FL 32111, USA}
\altaffiltext{22}{ELTE Gothard Astrophysical Observatory, H-9704 Szombathely, Szent Imre herceg st. 112, Hungary} 
\altaffiltext{23}{Department of Astronomy, University of Michigan, 830 Dennison, 500 Church St., Ann Arbor, MI 48109-1042, USA}
\altaffiltext{24}{Lehigh University, Department of Physics, 16 Memorial Drive East, Bethlehem, PA 18015, USA}
\altaffiltext{25}{Astrophysics Research Institute, IC2, Liverpool Science Park, Liverpool John Moores University, 146 Brownlow Hill, Liverpool, L3 5RF, UK}
\altaffiltext{26}{Department of Physics \& Astronomy, Johns Hopkins University, Baltimore, MD 21218, USA}
\altaffiltext{27}{NSF Astronomy and Astrophysics Postdoctoral Fellow}

\begin{abstract}
The \emph{Kepler} mission has yielded a large number of planet candidates from among the Kepler Objects of Interest (KOIs), but spectroscopic follow-up of these relatively faint stars is a serious bottleneck in confirming and characterizing these systems. We present motivation and survey design for an ongoing project with the SDSS-III multiplexed APOGEE near-infrared spectrograph to monitor hundreds of KOI host stars.  We report some of our first results using representative targets from our sample, which include current planet candidates that we find to be false positives, as well as candidates listed as false positives that we do not find to be spectroscopic binaries.  With this survey, KOI hosts are observed over $\sim 20$ epochs at a radial velocity precision of $100-200 \; \rm{m \; s^{-1}}$.  These observations can easily identify a majority of false positives caused by physically-associated stellar or substellar binaries, and in many cases, fully characterize their orbits. We demonstrate that APOGEE is capable of achieving RV precision at the $100-200 \; \rm{m \; s^{-1}}$ level over long time baselines, and that APOGEE's multiplexing capability makes it substantially more efficient at identifying false positives due to binaries than other single-object spectrographs working to confirm KOIs as planets. These APOGEE RVs enable ancillary science projects, such as studies of fundamental stellar astrophysics or intrinsically rare substellar companions. The coadded APOGEE spectra can be used to derive stellar properties ($T_{\rm eff}$, $\log g$) and chemical abundances of over a dozen elements to probe correlations of planet properties with individual elemental abundances.
\end{abstract}

\section{INTRODUCTION}
\subsection{\emph{Kepler's} Planet Candidates}
The \emph{Kepler} spacecraft's primary mission is to determine the frequency of Earth-sized exoplanets orbiting in the habitable zone of their parent stars \citep{bor2010,koc2010}, with a second objective of studying a wide variety of stellar astrophysics via asteroseismology \citep[e.g., ][]{cha2011}.  In addition, the high precision photometry \citep[$\sim 80 \; \rm{ppm}$ over 6-hour timescales for the brightest ($K_p \lesssim 15$) dwarfs, ][]{cal2010,gil2011,chr2012} enables studies of giant exoplanets and a wide variety of variable stars.  Its photometric band $K_p$ covers 423 - 897 nm and is similar to, but broader than, a combined $V$ and $R$ band \citep{koc2010}.  To find exoplanets, \emph{Kepler} makes use of the transit method, which detects planet candidates by measuring the flux loss that occurs when a planet crosses the face of its parent star.  However, there are several sources of false positives that must be taken into account when analyzing these candidates, most notably:  grazing eclipsing binaries (EBs), EBs (including hierarchical triples) whose eclipse depths are diluted by another star through flux contamination, brown dwarfs or low mass stars that have radii comparable to giant exoplanets, and even larger exoplanets that transit a fainter star within the photometric aperture.

Because of these sources of false positives, the \emph{Kepler} team makes a very clear distinction between candidate exoplanets and those that have been dynamically confirmed through spectroscopic radial velocity (RV) measurements or through photodynamical modeling \citep[e.g., ][]{hol2010,car2011}.  Kepler Objects of Interest (KOIs) consist of candidate exoplanets, eclipsing binaries, and known false positives.  Those KOIs that are not known to be false positives or EBs are referred to as ``active planet candidates'' \citep{bor2011a,bor2011b,bat2013}, but for simplicity, we will refer to such \emph{Kepler} planet candidates as ``KPCs'' throughout the rest of this paper.  An intermediate level of classification consists of ``validated'' exoplanets, which have very low probabilities of being blended EBs as determined through a Monte Carlo statistical analysis of the \emph{Kepler} photometry \citep[e.g., ][]{tor2011}.

\setcounter{footnote}{0}

As of October 2014, there are a total of 4229 KPCs amongst 3251 \emph{Kepler} stars
\footnote{\url{http://exoplanetarchive.ipac.caltech.edu/cgi-bin/ExoTables/nph-exotbls?dataset=cumulative_only}}
, but only $\sim 20$\% (653) of the stars host multiple KPCs.  It is estimated that as many as 15-26\% of transiting planets may have clearly detected transit timing variations \citep{ford2012}, which allow for mass determinations photometrically.  Even still, a majority of KPCs will require RV observations to confirm their planetary nature.  Such time-series RV observations are resource intensive, so efficient identification of false positive candidates is necessary to ensure efficient follow-up of likely planets.  In addition to aiding in the confirmation of KPCs, robustly determining the false positive rate amongst KPCs is required when conducting statistical analyses of this population.  A number of studies have attempted to perform such analyses, including investigations of planet frequency as functions of orbital periods and stellar host properties \citep{bor2011b,you2011,how2012}, and studies of the eccentricity distribution \citep{moo2011}.

Aside from the false positive rate of KPCs, knowledge of the host star(s) intrinsic properties (e.g., mass, radius, effective temperature, surface gravity, metallicity) is required to determine the masses and radii of the exoplanets,  as well as to conduct studies of planetary properties as functions of these stellar parameters.  The Kepler Input Catalog \citep[KIC, ][]{bro2011} provides a photometrically derived $T_{\rm{eff}}$, $\log{g}$, $\rm{[Fe/H]}$ and $\rm{E_{\left(B-V\right)}}$ for every star within \emph{Kepler}'s field of view through a combination of calibrated fluxes using $\left\{g, r, i, z\right\}$ filters similar to the original SDSS filters \citep{fuk1996} and a narrow-band \emph{D}51 filter modeled after the Dunlap Observatory \emph{DD}51 filter.  The catalog was originally used to inform target selection for the mission, but in the absence of a comprehensive spectroscopic survey of all $\sim 150,000$ \emph{Kepler} stars, the catalog's stellar parameters have been used in analyses of planet candidates.  There are ongoing efforts to provide improved stellar parameters of \emph{Kepler} targets by aggregating photometry, spectroscopy, asteroseismology, and transit analyses \citep{hub2014}.

\begin{figure}
\includegraphics[scale=0.4]{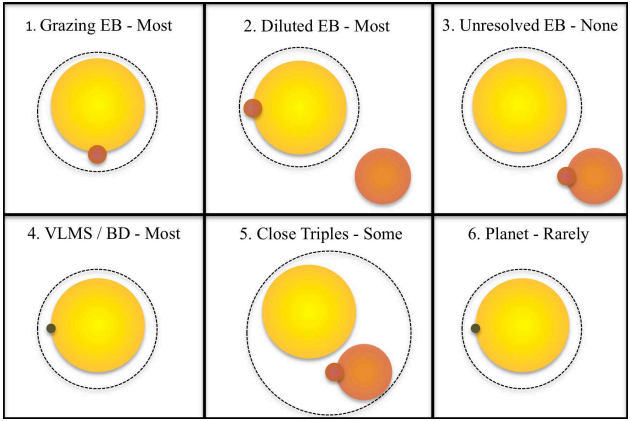}
\caption{The most common scenarios that can produce a lightcurve consistent with a transiting planet.  The titles qualitatively identify those scenarios that can be detected by a RV survey at the level of $\sim 100 \; \rm{m \; s^{-1}}$ (``Most'', ``Some'', etc.).  In each panel, the larger (yellow) star is the assumed KPC host star, while the dashed circles represent a spectrograph's FoV (not to scale).  All sources in each panel lie within the \emph{Kepler} photometric aperture, which is typically many arcseconds in size.  In Scenario 3, the term ``unresolved'' refers to the fact that the EB is unresolved in the \emph{Kepler} aperture.  In Scenario 6, only short-period, massive planets would be detected with APOGEE.  Note that giant planets can also be diluted to look like smaller-sized planets in these scenarios. \label{falseposdiag}}
\end{figure}

The majority of false positive KPCs are expected to be caused by astrophysical sources rather than random or systematic errors, specifically, EBs whose eclipse depths are similar to that expected from a transiting planet \citep{bor2011b}.  Fig.\ \ref{falseposdiag} demonstrates six of the most common sources of transiting KPC scenarios.  In each panel, the larger (yellow) star is the suspected KPC host, and all objects within the panels are assumed to be within the aperture used to create the \emph{Kepler} lightcurve.  Each \emph{Kepler} ``optimal aperture'' is variable, but is typically many arcseconds in size \citep{twi2010}.  The dashed circles represent a spectrograph fiber's field-of-view (FoV, not to scale).  The titles in each panel also denote, qualitatively, how often the given scenarios can be characterized by time-series RVs at modest precision ($\sim 100 \; \rm{m \; s^{-1}}$ level).  Note that in addition to stellar eclipses being diluted to look like giant planets, transits of giant planets can also be diluted to look like smaller planets.

\subsection{Sources Of False Positives}

Scenarios 1, 2, 4 and 6 all involve a physical companion orbiting the KPC host star.  In these scenarios, RV observations can detect the presence of grazing EBs (Scenario 1), EBs that are diluted by light from a third star within the \emph{Kepler} aperture, but resolved on-sky with the spectrograph (Scenario 2), or consist of a very low-mass star (VLMS) or brown dwarf companion (Scenario 4), a majority of the time.  An important sub-category of Scenarios 1 and 4 include EBs whose orbits produce only a secondary eclipse and no primary eclipse \citep{san2013}.  These false positives may be more common for longer period KPCs, where a companion star in an eccentric orbit is more likely to undergo secondary eclipse near periastron, but exhibit no primary eclipse.  In addition, the most massive, bona fide planets at short orbital periods will induce a Doppler velocity shift detectable at the $\sim 100 \; \rm{m \; s^{-1}}$ level (Scenario 6).  In those rare cases, their planetary nature will be confirmed through the APOGEE RVs by phasing them to the \emph{Kepler}-derived orbital period.

Stellar systems that are either physical multiples or visual companions with small separations on-sky, and consist of an EB, are represented by Scenario 5.  In this scenario, the diluted EB is only detectable if the flux of at least one component of the EB pair is sufficiently high that it appears in the cross-correlation function, or if the combined mass of the EB pair induces a sufficient velocity shift on the third (brightest) star in the system.  When the EB pair is composed of cooler K or M dwarf stars in the presence of a hotter primary, they are easier to detect in the NIR than in the optical, since the flux contrast is reduced in the $H$ band \citep[e.g., Kepler-16, ][]{ben2012}.  For a binary system composed of dwarf stars at a signal-to-noise of 100, secondaries with mass ratios down to $\sim 0.1$ are detectable in the $H$ band \citep{ben2008}, while mass ratios are limited to $\sim 0.5$ in the optical.  The detection limit for a given system depends on the number of stellar components within the aperture (e.g., is it a binary versus a triple system?), and whether any of those components are evolved (observing in the NIR is beneficial for components of differing $T_{\rm{eff}}$ ratios, not for brightness differences due to differing radii).

Scenario 3 represents a \emph{Kepler}-unresolved EB, where the variable star is within the \emph{Kepler} aperture, but is exterior to the spectrograph's FoV relative to the KPC host star.  This is the only scenario where RV observations will be not be able to detect any false positives, unless the RV survey targets every star within a given KPC's \emph{Kepler} aperture.  Fortunately, Scenarios 2, 3 and 5 can sometimes be tested photometrically with time-series photometry from the ground at greater spatial resolutions \citep[e.g., ][]{col2012}.  In addition, these are also the scenarios that are more likely to be solved using \emph{Kepler} data alone, e.g., by searching for flux centroid shifts.

\subsection{Paper Outline}
In this paper, we introduce our program to observe hundreds of KPCs using the Apache Point Observatory Galactic Evolution Experiment \citep[APOGEE,][]{maj2010} spectrograph \citep{wil2010, wil2012} on the Sloan 2.5 m telescope \citep{gun2006}, recently finished as part of the Sloan Digital Sky Survey III \citep[SDSS-III, ][]{eis2011} and continuing in SDSS-IV (2014-2020).  Our program provides an efficient means of determining the false positive rate of KPCs due to physically-associated binary stellar systems.  At the same time, these spectra are used in a variety of projects concerning the false positives themselves, including characterization of the orbits and measurements of the mass ratios for many of the spectroscopic binaries (SBs), or orbital characterization of intrinsically rare, massive ($M \gtrsim 10 \; M_{\rm{Jup}}$), substellar companions such as brown dwarfs and massive gas giant planets \citep{mar2000,sah2011}.  For KPCs that remain viable, host star properties such as $T_{\rm{eff}}$, $\log{g}$, and chemical abundances for dozens of elements can be derived using the APOGEE spectra.

In Section \ref{apgsection} we describe the APOGEE instrument and main survey, the methods used to derive RVs from its spectra, and its current RV precision floor.  In Section \ref{casestudies} we present RVs of five current and former KOIs observed during SDSS-III.  Three of these happened to be observed as part of a separate APOGEE EB program, and we present some conclusions on the nature of those KOIs as a precursor to our larger KPC campaign.  We also present the first results from our APOGEE-Kepler KOI campaign, using the (since confirmed) exoplanet host KIC 6448890 to test our long-term RV precision, and definitively identifying KIC 6867766 as a false positive exoplanet.

In Section \ref{qfacsection} we compare the efficiency of a survey using a high resolution, NIR, multi-object spectrograph against other planet-hunting spectrographs: HARPS-North, which is a clone of HARPS-South with some improvements \citep{may2003}, Keck HIRES \citep{vog1994}, SOPHIE \citep{per2008}, and HET HRS \citep{tul1998}.  We demonstrate that by using a multiplexing instrument in the NIR to conduct a survey at modest RV precision ($100 \; \rm{m \; s^{-1}}$), false positives can be identified more efficiently compared to the single-object instruments, reserving telescope time on those other resources for confirmation of the remaining KPCs at significantly higher precision.  In Section \ref{fapsection} we review other techniques for determining the false positive rate of \emph{Kepler} KPCs, and highlight the science enabled by extracting abundances from the coadded APOGEE spectra.  We summarize our findings in Section \ref{summarysection}.

\section{APOGEE Survey Overview}
\label{apgsection}
APOGEE is a survey of Milky Way stars using a multi-object, fiber-fed, NIR spectrograph housed in a vacuum cryostat, that can observe up to 300 objects simultaneously, producing $R \sim 22500$ spectra covering a wavelength range of $1.51 - 1.68 \; \rm{{\mu}m}$ using a volume phase holographic grating mosaic.  Details of the instrument design can be found in \citet{wil2010} and \citet{wil2012}.  Typically the instrument achieves a signal-to-noise ratio per pixel of 100 ($\Delta \lambda \sim 0.1 - 0.17 \AA$) on an $H = 11$ star in a single visit (one hour of total integration).  Most stars are observed on a minimum of three different nights, so that short-period binaries can be flagged.  Each field on the sky is normally observed in multiples of three, ranging from 3 to 24 epochs, with brighter targets swapped for new stars after three observations.  Aluminum plug plates hold optical fibers that carry the star light from the telescope into the instrument.  The primary science goal of the survey is to study the Milky Way by measuring radial velocities and chemical abundances of $\sim 10^5$ red giant stars, but a variety of additional science projects are included.  A summary of the project can be found in \citet{all2008b}.  A detailed description of the survey will appear in \citet{maj2015}.  Details of the target selection for the survey in SDSS-III can be found in \citet{zas2013}.

\begin{figure}
\includegraphics[scale=0.5]{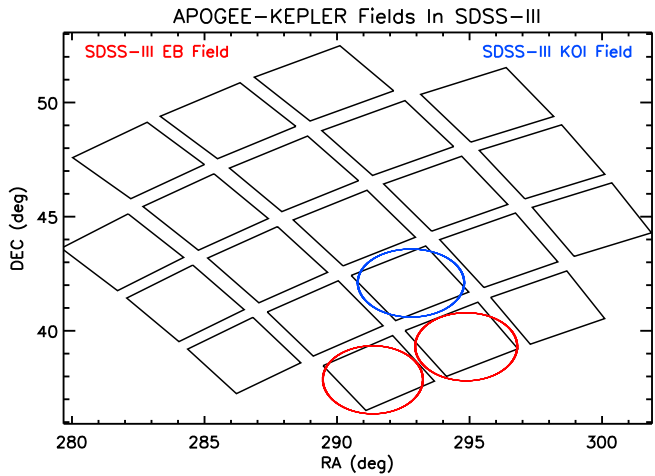}
\caption{\emph{Kepler} module footprints with APOGEE FoV (circles, $2.98^\circ$ diameters) from our SDSS-III \emph{Kepler} EB and KOI programs.  Five additional \emph{Kepler} modules will be observed by APOGEE in SDSS-IV.\label{apgkepfov}}
\end{figure}

The telescope's field-of-view (FoV) covers a circular area $1.49^{\circ}$ in radius, which matches well to the size of a given \emph{Kepler} module.  Fig.\ \ref{apgkepfov} shows the \emph{Kepler} modules' footprints, along with the three SDSS FoVs for our programs observed during SDSS-III.  A total of 163 KPC host stars are observed in the SDSS-III KOI field (blue).  Those targets were selected from all KPCs that had $H < 14$, 153 of which were dispositioned as planet ``candidates'' as of August 2013 (four others were confirmed exoplanets, six were not dispositioned yet).  As can be seen, a single SDSS footprint covers most of a \emph{Kepler} module's FoV.  In addition to the KPC hosts observed during SDSS-III, five additional \emph{Kepler} modules will be observed in SDSS-IV.

The APOGEE data processing pipeline is described in \citet{nid2015}.  Basic steps include collapsing the detector exposures for each of the three NIR arrays from 3D data cubes to 2D images, flat fielding, aperture extraction, wavelength calibration, sky subtraction, telluric correction, and measurement of RVs.  The RVs currently calculated by the automated pipeline make use of a grid of synthetic spectra calculated from ATLAS9 stellar model atmospheres \citep{mes2012,zam2015}.  The mean, internal RV precision of the pipeline-produced relative RVs is $\sim 100 \; \rm{m \; s^{-1}}$, although it does depend on signal-to-noise, spectral type, and level of residual systematics from the data processing.  Critically, these pipeline-derived RVs only work for simple cross-correlation functions, and are expected to fail when there is contamination from multiple stellar spectra, as in the case for most binary stars.  As such, we derive our own RVs using additional, interactive processing of the data.  These steps include manually correcting residual OH sky emission lines, selecting templates from a finer grid, and interactively fitting cross-correlation peaks, which may often be asymmetric or have multiple components in the case of binary stars.  We calculate uncertainties for our RV measurements following the maximum-likelihood procedure laid out by \citet{zuc2003}.  This approach derives an analytical relationship between the cross-correlation function and it's first and second derivatives to account for uncertainty contributions related to the sampling and sharpness of the correlation peak, and the signal-to-noise of the target and template spectra.  The RVs are then fit using a custom wrapper to the RVLIN software package \citep{wri2009}, which includes the ability to fit both components of a double-line spectroscopic binary through an iterative approach, and forces some orbital parameters to be identical between both components (e.g., orbital period, eccentricity, epoch of periastron).  In some cases we make use of our IDL-based Levenberg-Marquardt fitting code used in \citet{ben2012}.

\section{Results}
\subsection{Initial Case Studies}
\label{casestudies}
We have selected five KOIs with diverse histories and current statuses to test and develop our analysis pipelines.  Three of these targets were observed as part of an SDSS-III ancillary program studying \emph{Kepler} EBs \citep{mah2015} (hereafter MAH2015).  These targets were at one point \emph{Kepler} planet candidates, but were determined to be likely EBs by the time the MAH2015 observations began.  Note that since these KOI hosts were observed through a different program, the total number of epochs for these targets is less than the number of epochs that the SDSS-Kepler KOI program obtains (the SDSS-III EB program obtained $3-6$ epochs for each target, compared to $>18$ epochs for the SDSS-III KOI program).  The two other KPC hosts presented here come from our SDSS-III Kepler KOI program.  We summarize our findings on these targets to demonstrate the diversity of astrophysical configurations encountered in our spectroscopic observations.

\subsubsection{SDSS-III EB Program: KIC 1571511}
KIC 01571511 (KOI 362, $K_p = 13.42$, $H = 12.04$), consists of an F-type dwarf and low-mass M dwarf \citep{ofi2012}, and was observed as part of MAH2015.  The orbital period is 14.0224519 days and the radius as estimated in the NExScI KOI catalog\footnote{\url{http://exoplanetarchive.ipac.caltech.edu}} is $14.7\;\pm\;6.4 \; R_\oplus$.  The eclipses of such a low-mass star ($\sim 2$\% decrease in flux during primary eclipse) are comparable to those expected for a gas giant planet.  Indeed, this star was originally suspected to be an overlooked gas giant exoplanet \citep{cou2011,ofi2012}.  In the specific case of KIC 01571511, there is a small secondary eclipse ($\sim 0.05$\%) detected in the \emph{Kepler} lightcurve, which can be used to derive an estimate of the relative $T_{\rm{eff}}$ ratio between the primary and secondary, and can therefore be used to help determine whether the object is a likely stellar companion.  However, there is no guarantee that an EB system with a primary eclipse will also show a secondary eclipse, nor that the secondary eclipse is detectable even with \emph{Kepler's} precision.  In fact, \citet{san2012} found that some of their false positive KPCs were EBs in eccentric orbits for which only the shallower, secondary eclipses are present, but were mistaken as planetary transits across the primary.

\begin{figure}
\includegraphics[scale=0.5]{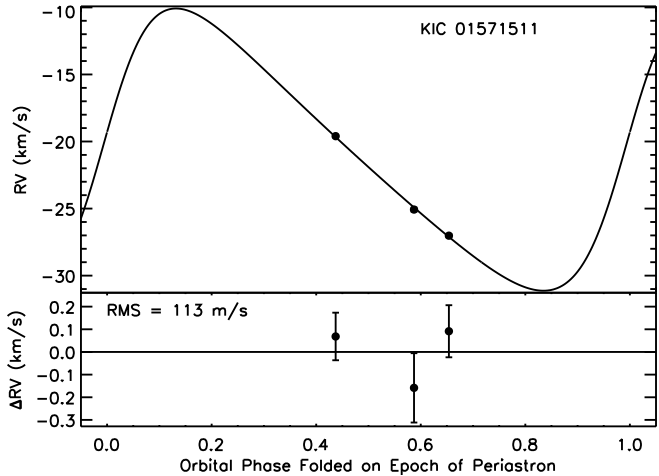}
\caption{APOGEE RVs of the \emph{Kepler} EB KIC 01571511 and the best-fit model from \citet{ofi2012} shown as the solid line.  A constant offset between the model and the APOGEE RVs has been applied to account for a zero-point offset.  For the given \emph{Kepler} period and epoch of transit, it is clear even with just three APOGEE RVs that the object is not a planet, because the change in RV over a short fraction of the orbit is much greater ($\sim 10 \; \rm{km \; s^{-1}}$) than expected for a planetary mass.\label{apgofir}}
\end{figure}

Fortunately, these EBs are fairly trivial to detect spectroscopically, as demonstrated in Fig.\ \ref{apgofir}.  In this figure, we plot the best-fit RV model from the analysis by \citet{ofi2012}, noting that there is a typo in the value of $\omega_1$ in their Table 3 that is missing a minus sign.  We also plot the three APOGEE RVs obtained through the ancillary program (Table \ref{apg_rv_mastertable}).  Only a constant offset between the model and APOGEE data is included to account for instrumental zero-point differences.  Even with three data points, the RV variation observed in the APOGEE RVs is inconsistent with a giant planet, given the period and epoch of transit from the \emph{Kepler} lightcurve, because the change in RV over a short fraction of the orbit is much greater ($\sim 10 \; \rm{km \; s^{-1}}$) than expected for a planetary mass.  These data also demonstrate that APOGEE is capable of producing RVs at the $\sim 100-200 \; \rm{m \; s^{-1}}$ for $H \sim 12$ stars based on the rms residual to the well-determined orbital solution from \citet{ofi2012}.  KIC 1571511 corresponds to Scenario 4 in Fig.\ \ref{falseposdiag}, where a low-mass star generates an eclipse depth comparable to that expected from a giant planet.

\subsubsection{SDSS-III EB Program: KIC 3848972}
KIC 3848972 (KOI 1187, $K_p = 14.49$, $H = 12.80$) is listed in both the EB and KOI catalogs, and was observed as part of MAH2015.  As a KOI, the target was listed as a ``False Positive'' in the Q1-Q8 catalog, but is currently absent in the Q1-Q16 catalog.  The KOI Q1-Q8 catalog lists a period of $P = 0.37052915$ days, while the EB catalog lists a period that is twice as long ($P = 0.741057$ days).  The estimated radius reported in the KOI catalog is $3.53\;\pm\;0.93 \; R_\oplus$.  Where the EB Catalog assumes two nearly-equal eclipses from primary and secondary eclipse events, the KOI catalog reports half the orbital period and defines the secondary eclipse as undetected or absent.  Multi-color, ground-based photometry was observed by \citet{col2012} using a tunable filter on the OSIRIS instrument on the 10.4-m Gran Telescopio Canarias (GTC).  They find a consistent star-planet radius ratio ($2\sigma$) in both their blue and red filters, but measure a statistically significant ($5.8\sigma$) difference in the eclipse depths.  Interestingly, the color differences during eclipse suggest that the secondary component is bluer than the primary.

\begin{figure}[t]
  \centerline{
    \includegraphics[width=4.5cm]{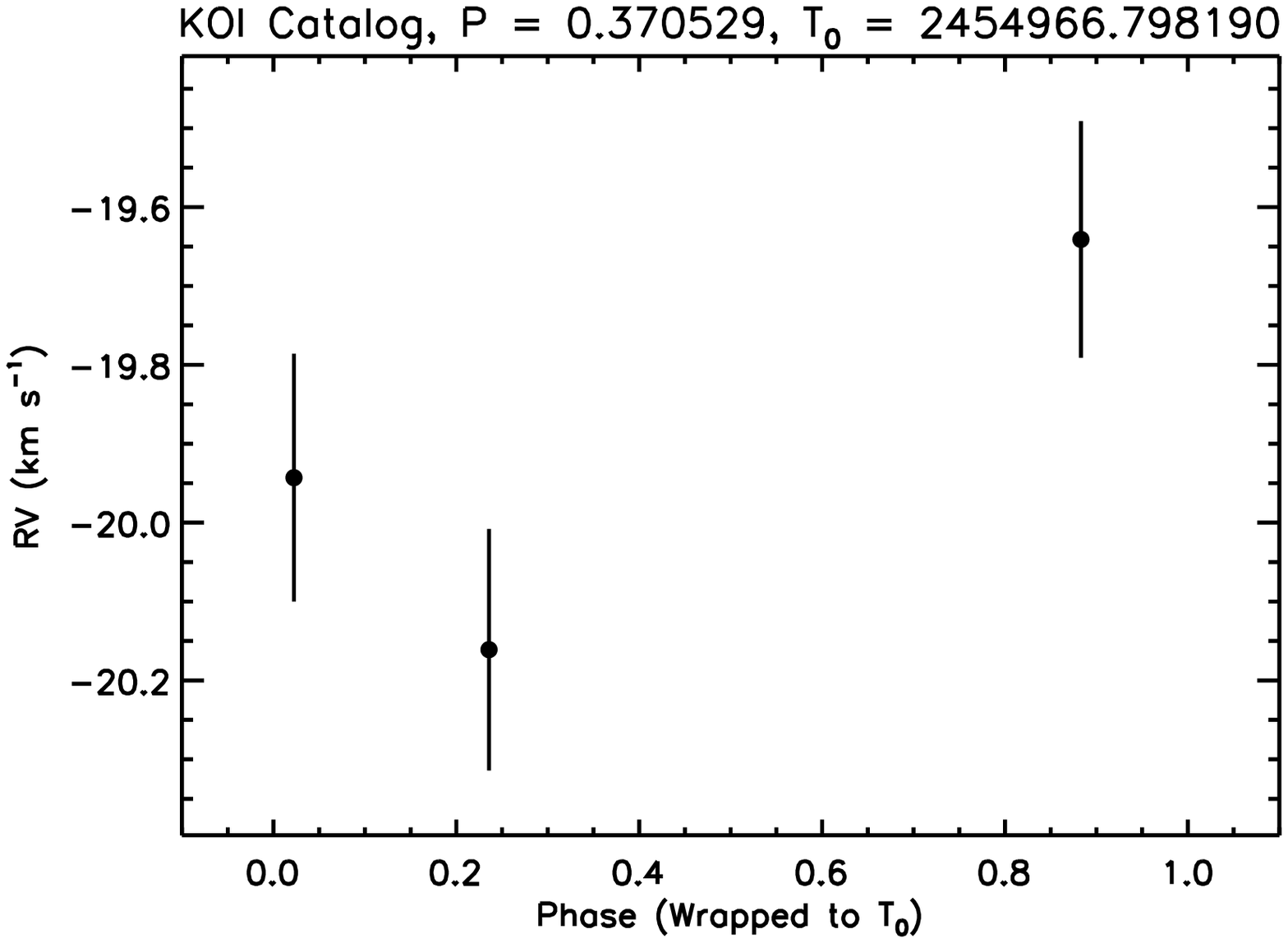}
    \includegraphics[width=4.5cm]{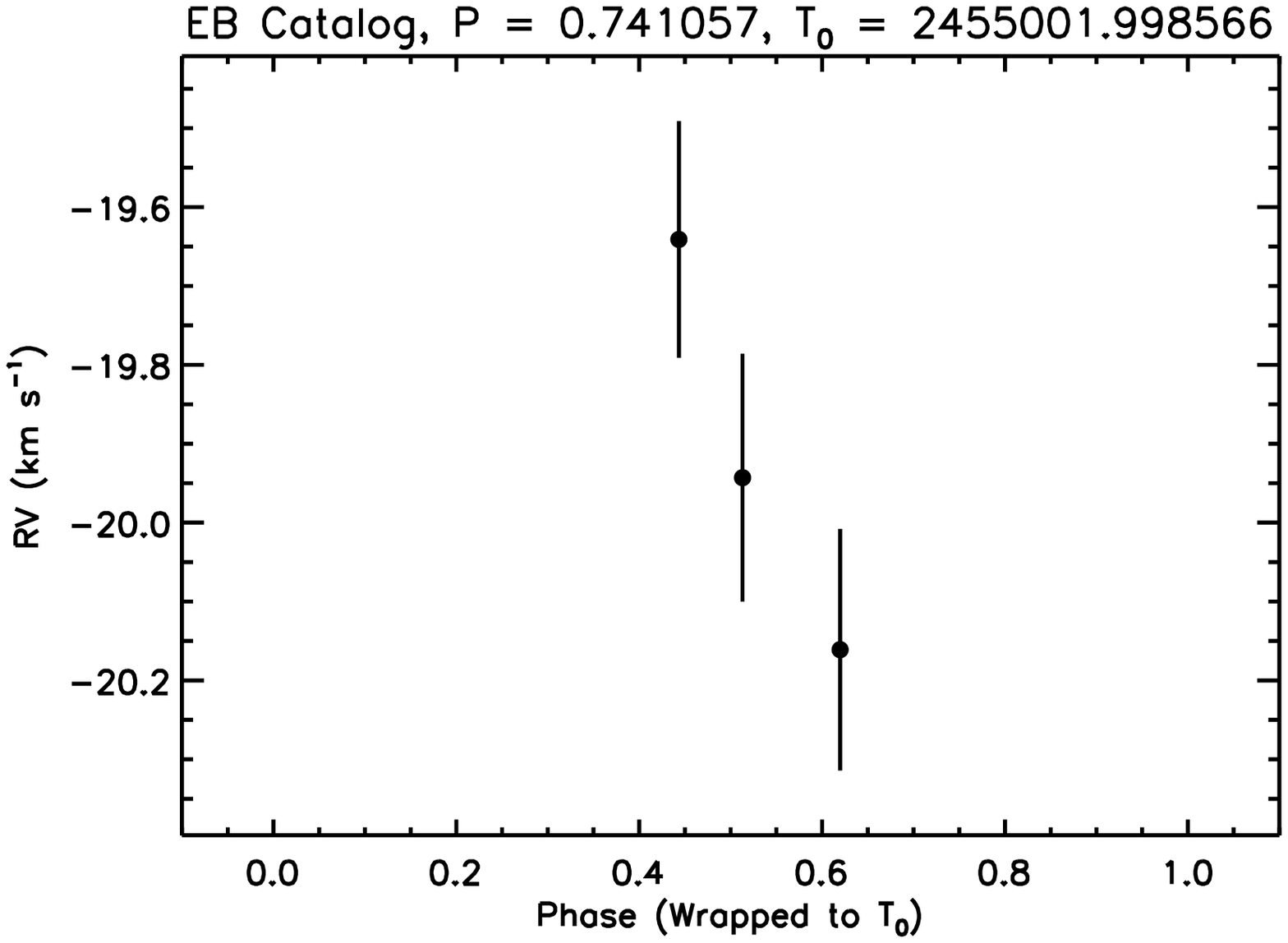}
  }
\caption{Orbital phase-folded RVs for KIC 3848972, using the period and ephemeris values from the KOI Q1-Q8 Catalog (left) and EB Catalog (right).  The period ambiguity arises from whether the system is treated as an object that only produces a primary transit (KOI solution), or an eclipsing binary that produces both a primary and secondary eclipse with similar depths (EB solution).  No significant RV variation is seen beyond a few hundred $\rm{m \; s^{-1}}$, disfavoring a physically bound stellar companion at either of these orbital periods as the source of the \emph{Kepler} signal.\label{kic3848972}}
\end{figure}

Only three APOGEE spectra were obtained for this target as part of MAH2015, however, a check on binarity can still be performed provided the orbital phase coverage is reasonable.  We conduct a one-dimensional cross-correlation using a K-type dwarf template ($T_{\rm{eff}} = 5000 \rm{K}$).  We do not see evidence for any significant rotational broadening greater than $\sim 10-20 \; \rm{km \; s^{-1}}$.  Although the APOGEE spectra for this star are somewhat noisy, we find a single, very stable CCF peak with no RV variation greater than a few hundred $\rm{m \; s^{-1}}$ (Table \ref{apg_rv_mastertable}).  There is no obvious correlation with orbital phase after folding on both the KOI and EB Catalog periods and ephemerides, despite spanning $\sim 80$\% of the KOI orbital phase and $\sim 20$\% of the EB Catalog orbital phase, respectively (Fig.\ \ref{kic3848972}).  If the signal was caused by a hotter (bluer) secondary orbiting a brighter primary, the expected RV amplitude should be many tens of $\rm{km \; s^{-1}}$.

Another possible explanation is that the observed \emph{Kepler} signal comes from an object transiting a low-mass, cool (red) star that is within the photometric aperture (constrained to be $r_a < 2''$ given the \citet{col2012} GTC aperture), but too faint to detect spectroscopically with APOGEE in the presence of the bright primary star.  In this case, the color-dependent transit depths are caused because the fainter, redder component is the one being transited, hence the overall color of the combined light appears to shift towards the blue during the transit event.  Intrigued by this possibility, we obtained Keck adaptive optics (AO) imaging to search for a fainter companion that might be the source of the \emph{Kepler} signal.

The AO image was acquired on UT 2014 Jul 17, using NIRC2 (instrument PI: Keith Matthews) and the Keck II Natural Guide Star (NGS) AO system \citep{wiz2000}.  We used the narrow camera setting with a plate scale of $10 \; \textrm{mas pixel}^{-1}$, which provides a fine spatial sampling of the instrument point spread function (PSF).  The observing conditions were excellent, with a seeing of 0$^{\prime\prime}.3$.  KIC 3848972 was observed at an airmass of 1.12.  We used the $K_S$ filter to acquire the image using a 3-point dither method.  At each dither position, we took a total of 10 coadds composed of 5-second exposures. The total on-source integration time is therefore 150 seconds.

The raw NIRC2 data were processed using standard techniques to replace bad pixels, flat-field, subtract thermal background, align, and coadd frames.  We calculated the $5\sigma$ detection limit as follows.  We first defined a series of concentric annuli centered on the star.  For the concentric annuli, we calculated the median and standard deviation of flux for pixels within these annuli.  We define the $5\sigma$ detection limit as five times the standard deviation above the median flux.  Representative $5\sigma$ detection limits are \{1.6, 3.3, 5.0, 5.4\} magnitudes for projected separations of \{0$^{\prime\prime}.1$, 0$^{\prime\prime}.2$, 0$^{\prime\prime}.5$, 1$^{\prime\prime}$\}, respectively.

\begin{figure}
\includegraphics[scale=0.5]{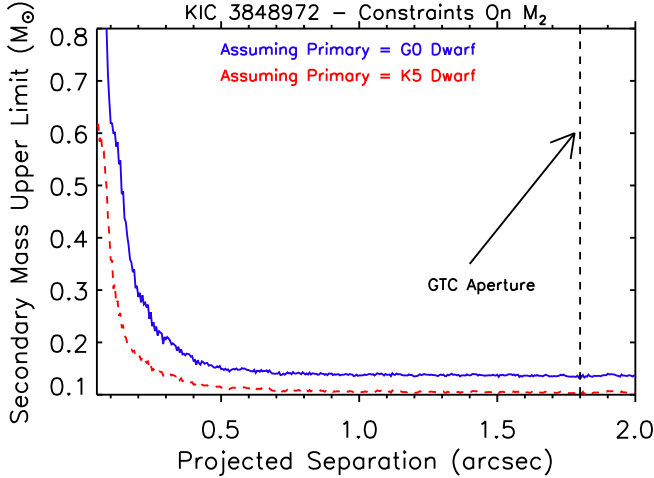}
 \caption{Upper limits on secondary companion mass from Keck AO imaging.  The photometric aperture used by \citet{col2012} is marked by the vertical line, and serves as the outer limit of where the eclipsing object might lie.\label{kic3848972_keckao}}
\end{figure}

We translate the $5\sigma$ upper limits on companion brightness into upper limits on companion mass using the SED models compiled in \citet{kra2007}, assuming the secondary is a bound companion and that differential extinction between the two spectral types is minimal in the $K_s$ band.  For a given absolute $K_s$ magnitude of the primary, the contrast curve from the Keck AO data gives a lower limit for the secondary's absolute $K_s$ magnitude, which we then interpolate into a mass using the \citet{kra2007} models.  We adopt primary spectral types of G0 and K5 as conservative upper and lower limits based on our spectroscopic cross-correlation analysis.  We are able to rule out any bound companions more massive than $0.2 \; M_{\odot}$ at the $5\sigma$ level exterior to 0.2 arcseconds (Fig.\ \ref{kic3848972_keckao}).

Alternatively, the transit signal could be caused by a fainter background or foreground star that is physically unassociated with KIC 3848972, but still within the \emph{Kepler} photometric aperture.  Following \citet{mor2011}, we can estimate the probability of having a blend source within a given aperture using a model of the galactic population from TRILEGAL\footnote{\url{http://stev.oapd.inaf.it/cgi-bin/trilegal_1.6}} \citep{gir2012}.  We generate a TRILEGAL population in a one square degree area centered on KIC 3848972 with the default settings (see Appendix \ref{appendix1}).  We calculate a mean stellar density of $0.004581 \; \rm{stars \; arcsec^{-2}}$ within \emph{Kepler} magnitudes $15.49 \leq K_p \leq 21.26$.  This magnitude range is chosen because it represents stars faint enough to be undetected in the \emph{Kepler} aperture but still able to produce a transit depth of $\delta = 0.00196$.  With this mean density, the probability of just finding a potential blend source within this magnitude range is 5.8\% within 2\arcsec (the GTC photometric aperture) and just 0.36\% within 0\farcs5 (the area least probed by the Keck AO images, Fig.\ \ref{kic3848972_keckao}).  From \citet{mor2011}, the probability that this blend source is an EB with a configuration that could mimic a transiting planet signal is on the order of 2.5E-4, so the probability of having a background EB as the source of this KPC is on the order of \{9E-7, 1.4E-5\} within \{0.5, 2.0\} arcseconds, respectively.  Thus, the more likely EB false positive scenario is a bound eclipsing binary causing the transit signal.

The lack of observed RV variability indicates this KOI is not due to a physically bound stellar companion orbiting the brightest component of the KIC 3848972 system, while the Keck AO images constrain any bound, diluted EB to be either within $\sim 0.5^{\prime\prime}$ of the primary, or more than 5 magnitudes fainter than the primary in the $K_s$ band.  Given the \citet{col2012} observations, we hypothesized that this KOI corresponded to Scenario 2 or 5 in Fig.\ \ref{falseposdiag}, but our observations rule out Scenario 2 and tightly constrain the separation of a diluted EB under Scenario 5.

\subsubsection{SDSS-III EB Program: KIC 3861595}
KIC 3861595 (KOI 4, $K_p = 11.43$, $H = 10.27$) is listed in both the EB and KOI catalogs, and was observed as part of MAH2015.  As a KOI, the target was initially listed as a ``False Positive'' in the Q1-Q8 catalog, and is currently listed as ``Not Dispositioned'' in the Q1-Q16 catalog.  The orbital period is 3.8493724 days and the estimated planet radius is $11.8\;\pm\;1.6 \; R_\oplus$.  Some ground-based observations have been conducted and reported at the Kepler Community Follow-up Program (CFOP) website\footnote{\url{https://cfop.ipac.caltech.edu/home/}}.  These include several optical spectra from the TRES spectrograph \citep{sze2007} that indicated the star was a rapid rotator (40-50 $\rm{km \; s^{-1}}$), and potentially variable at a level of a few hundreds of $\rm{m \; s^{-1}}$.  Imaging from the 1-m Nickel telescope at Lick Observatory and Keck HIRES guider images show two nearby stars within ten arcseconds of the target.  Both nearby stars appear to be approximately 6 magnitudes fainter than the target.

In addition to six APOGEE spectra, MAH2015 obtained five optical spectra for this target using the High Resolution Spectrograph (HRS) on the Hobby-Eberly Telescope.  The HRS was used in the 30,000 resolution mode, with the 316 $\rm{g/mm}$ grating at a central wavelength of $\lambda_0 = 5936 \; \AA$.  The HET spectra were reduced using our optimal extraction pipeline described in MAH2015.  We find a good template match (for both HET and APOGEE) using a mid-F spectral template rotationally broadened to 40 $\rm{km \; s^{-1}}$, in agreement with the CFOP notes from the TRES observations.  The estimated spectroscopic rotation rate of $\sim \; 40 \; \rm{km \; s^{-1}}$, combined with an estimated rotation rate of 5.65-5.8 days \citep{hir2012,rho2014}, results in an equatorial radius of $\sim \; 4.5 \; R_{\odot}$, somewhat larger than the spectroscopically determined radius of $2.727 \; \pm \; 0.504 \; R_{\odot}$ \citep{buc2012} and the Stellar Parameter Catalog's value of $2.992 ^{+0.469} _{-0.743} \; R_{\odot}$ \citep{hub2014}, which also uses the spectroscopic stellar parameters of \citet{buc2012}, but tie the stellar parameters to Dartmouth stellar evolution models \citep{dot2008}.

We cross-correlate the HET and APOGEE spectra with mid-F spectral templates rotationally broadened to 40 $\rm{km \; s^{-1}}$.  For this target, we used subsections of the APOGEE spectrum (1.515-1.560 ${\mu}\rm{m}$, 1.586-1.605 ${\mu}\rm{m}$, 1.615-1.635 ${\mu}\rm{m}$, and 1.6475-1.6775 ${\mu}\rm{m}$) to avoid several of the broadest lines.  We find a single-peaked, broad cross-correlation function, with RV variation at the $\sim \; 1 \; \rm{km \; s^{-1}}$ level (Table \ref{apg_rv_mastertable}).  We note that the RV scatter is larger than most of the A and F stars observed by \citet{lag2009}.  However, phase-folding and fitting both sets of RVs at the orbital period and ephemerides found in the KOI and EB catalogs fails to resolve a signal of orbital motion from a bound companion at that period.

\begin{figure}
\includegraphics[scale=0.5]{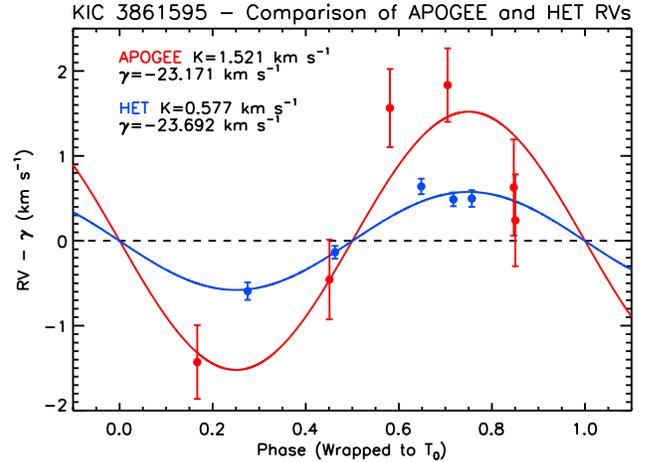}
\caption{Phase-folded RVs for APOGEE (red) and HET (blue).  Each set were fit with the orbital period and transit ephemeris fixed to the KOI value.  Eccentricity is forced to zero.  We find that while both sets of RVs appear to be in-phase with the orbital parameters, the RV semiamplitudes are quite different.  This suggests the spectrum might be blended (Scenarios 2 or 5 in Fig.\ \ref{falseposdiag}): the APOGEE spectra can be more sensitive to such a blend if the temperatures of the blended components are different.  Line bisector variations also suggest a blend, but is not definitive due to the small number of observations.  It is also possible that the uncertainties are underestimated due to the rapid rotation of the star ($40 \; \rm{km \; s^{-1}}$). \label{kic3861595}}
\end{figure}

\begin{figure}
\includegraphics[scale=0.35]{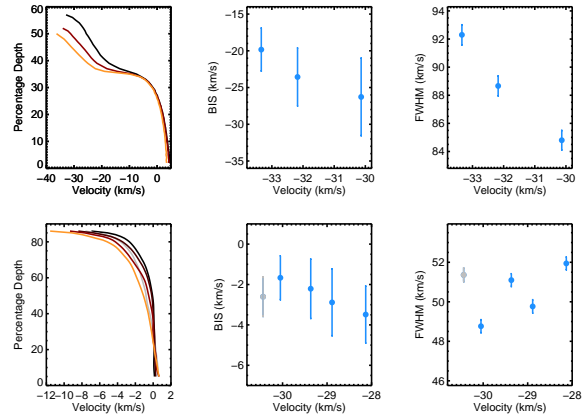}
\caption{Bisector analysis for the three APOGEE spectra that were on a common fiber (top row) and HET spectra (bottom row). The grey data point in the HET plots represents a low signal-to-noise observation. (Left: ) Bisectors of the cross-correlation function, shifted by measured radial velocities. Colors are based on bisector inverse slope (BIS) values. Percentage depth is a proxy for flux, and does not span the full range from 0 to 1 because of continuum ambiguities. (Middle: ) Correlation between BIS and measured RV. (Right: ) Correlation between FWHM of the CCF, and measured RV. Uncertainties are determined from the measurement variation between echelle orders (for HET spectra) and between the three detectors (for APOGEE spectra). Note that difficulty in defining the CCF continuum probably leads to an overestimation of the BIS errors. \label{kic3861595_bisectors}}
\end{figure}

In fact, we find that if we fit each set of RVs separately, fix the period and ephemeris to the KOI values, and force the eccentricity to zero, the best-fit RV semiamplitudes differ by a factor of 2.5 (Fig. \ref{kic3861595}).  A color-dependent semiamplitude may signal a blended spectrum (Scenarios 2 or 5); the redder component can affect the line shapes more significantly in the NIR, but such scenarios are particularly challenging to identify in rapidly rotating stars using only a handful of observations.  Nevertheless, we undertake a full line bisector analysis of both HET and APOGEE spectra. After creating custom numerical stellar template masks for both the HET and APOGEE wavelength ranges, we calculate the bisectors of the cross-correlation function, similar to the procedure described in \citet{wri2013}. We are limited to using three of the six APOGEE observations because they were the only ones observed on the same plug fiber, and therefore should have the same intrinsic profile. The bisectors appear to be varying both in shape and position (indicating a cause other than bulk motion of the primary) and the bisector inverse slope (BIS) seems well-correlated with both RV and CCF FWHM (Fig.\ \ref{kic3861595_bisectors}). However, the rapidly-rotating nature of this star causes difficulty in establishing the CCF continuum, and complicates bisector analysis. Any attempt to calculate errors on the BIS leads to overestimation, and therefore we are hesitant to quantify this result beyond saying that a blend scenario is possible.

We can not definitively show the transit signal is caused by a spectroscopic blend with so few bisector measurements.  Our HET RVs are consistent with a planetary companion in a circular orbit, but there are not enough RVs to make a firm claim.  The APOGEE RVs contradict this claim, but RV uncertainties for rapid rotators have not been thoroughly vetted, so the APOGEE RV uncertainties reported in Table \ref{apg_rv_mastertable} could be underestimated.  We also note that an analysis by \citet{rho2014} found that an Algol-type background binary, approximately 6.5 magnitudes fainter but within the \emph{Kepler} photometric aperture, could produce a lightcurve similar to what's observed.  This scenario, which corresponds to Scenario 3 in Fig.\ \ref{falseposdiag}, also remains a possibility given the companions seen in the Lick and Keck images at this approximate flux ratio.  This KOI is a prime example as to why it is sometimes necessary to obtain multiple spectra when searching for exoplanet false positives; obtaining just a few spectra, even at orbital quadratures, may not be sufficient to confidently identify a blended stellar binary, especially for systems that are rapid rotators.  Additional spectroscopic observations will be able to study the line bisectors and RVs in sufficient detail to determine the nature of this intriguing KOI.

\subsubsection{SDSS-III KOI Program: KIC 6448890}
\begin{figure}
\includegraphics[scale=0.5]{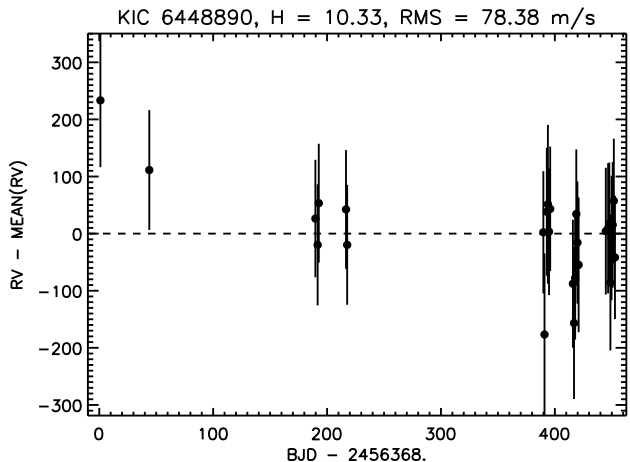}
\caption{The long-term RV rms of the confirmed exoplanet host star KIC 68448890 demonstrates we are able to achieve relative RV precision at the $100 \; \rm{m \; s^{-1}}$ level.\label{kic6448890_rvs}}
\end{figure}

KIC 6448890 (KOI 1241, $K_p = 12.44$, $H = 10.33$) is a system with two exoplanets that have been confirmed via transit timing variations \citep{ste2013}.  The two planets have orbital periods of 10.5016 and 21.40239 days, radii of 0.581 and 0.874 $R_{\rm{Jupiter}}$, and masses of 0.07 and 0.57 $M_{\rm{Jupiter}}$ \citep{hub2013}.  The RV semiamplitude is too small to be detectable with APOGEE, so this target ($H=10.33$) is an opportunity to test the long-term RV precision level for our KOI program.  Fig.\ \ref{kic6448890_rvs} shows that the RV rms about the mean is $78 \; \rm{m \; s^{-1}}$ over the entire baseline, in support of our stated goal to achieve a long-term (1-2 year), relative RV precision of $\sim 100 \; \rm{m \; s^{-1}}$.  The APOGEE RVs are reported in Table \ref{apg_rv_mastertable}.

\subsubsection{SDSS-III KOI Program: KIC 6867766}
KIC 6867766 (KOI 1798, $K_p = 14.38$, $H = 12.99$) was listed in the Q1-Q6 KOI catalog as an exoplanet candidate but has since been listed as ``Not Dispositioned'' in later catalogs.  This KOI is also listed in the EB catalog.  A shallow, 0.3\% transit signal is present with no obvious secondary feature at an orbital period of 12.964725 days.  The estimated radius from the KOI catalog is $9.65\;\pm\;1.5 \; R_\oplus$.  This target was observed a total of 25 times with APOGEE as part of our \emph{Kepler} KOI program within SDSS-III.  We found the best 1D CCF template match using a $T_{\rm{eff}} = 5500$ K, solar-metallicity dwarf rotationally broadened to 14 $\rm{km \; s^{-1}}$.  Upon visual inspection, some of the CCFs were observed to be asymmetric, and in some cases evidence of a double-peaked CCF were present.  In these situations, the parameters of our best 1D CCF template are generally not reflective of any component in the system: a multi-dimensional cross-correlation analysis is required.

\begin{figure}
\includegraphics[scale=0.5]{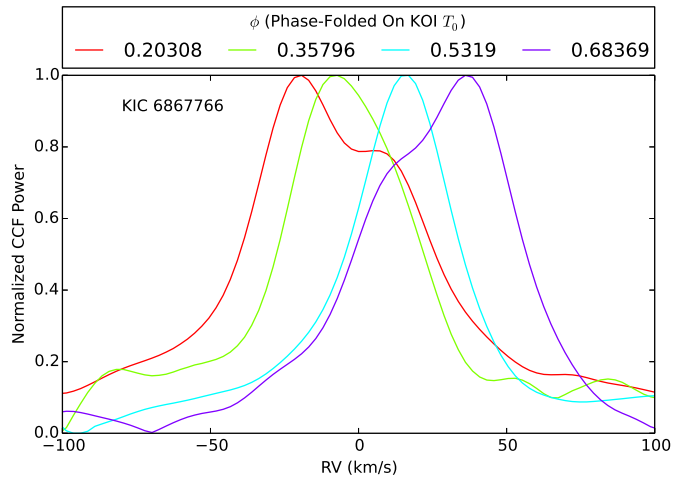}
\caption{Phase-folded 1D CCF for KIC 6867766 as observed by APOGEE for a subset of the epochs.  The CCF is clearly double-peaked, and the two CCF components have maximum separation near phases $\phi \sim 0.25$ and $\phi \sim 0.75$: a clear indication that the KOI planet signal is the result of a binary companion.\label{ccftimevol}}
\end{figure}

To test whether the \emph{Kepler} transits might be due to the binary showing up in the CCF, we calculate the 1D CCFs for each APOGEE observation, normalize each CCF to a peak value of unity, and then sort them based on the orbital phase corresponding to each observation, phase-folding on the time of transit and orbital period as reported in the KOI catalog.  We show only a few of the CCFs sampling different orbital phases in Fig.\ \ref{ccftimevol} for clarity.  We find that this technique is quite effective at finding time-variable CCF changes indicative of blended SB2s.  In the case of KIC 6867766, the double-peaked nature of the CCF is readily apparent, and phase-folds nicely with the KOI orbital solution (maximum separation between CCF components near phases $\phi \sim 0.25$ and $\phi \sim 0.75$ after phase-folding on KOI $T_0$, blended components near $\phi \sim 0$ and $\phi \sim 0.5$).

Upon further analysis, we determined that this system includes three stellar mass objects: a late F-type dwarf (KIC6867766A) with an early M-dwarf companion (KIC6867766B) in a $\sim 13$ day orbit, and a late G-type dwarf (KIC6867766C) with no discernible velocity motion.  We analyzed the APOGEE spectra with the TRICOR algorithm \citep{zuc1995}, which uses a 3D cross-correlation to derive the RV of three blended spectral components and their relative flux ratios.  Our analysis used spectral templates constructed from BT Settl models \citep{ala2011}, convolved to the APOGEE spectral resolution and sampling.  We optimized the template stellar parameters to best match the observed spectra, using the peak correlation power from our highest S/N APOGEE spectra to assess the template match.  All three components used templates with solar metallicity and $\log{g} = 4.5$.  The effective temperatures and rotational velocities were: A: 6200 K, $10 \; \rm{km \; s^{-1}}$; B: 3500 K, $3 \; \rm{km \; s^{-1}}$; C: 5200 K, $3 \; \rm{km \; s^{-1}}$.  We fixed the relative $H$ band flux ratios to be C/A = 0.35 and B/A = 0.05, which are consistent with both the optimal flux ratios derived by TRICOR and the physical flux ratios expected for these stars.  

\begin{figure}
\includegraphics[scale=0.5]{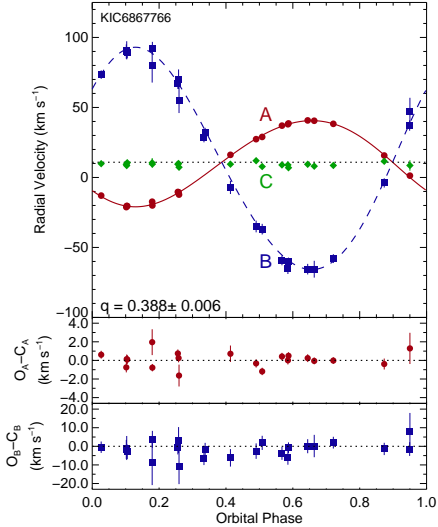}
\caption{Three-component RV solution for KIC 6867766 using APOGEE RVs.  Components A and B are an F+M EB pair that produces the \emph{Kepler} transit signal, while Component C is a (likely) bound G dwarf companion with a long orbital period that dilutes the A+B eclipse.\label{kic6867766_sb3}}
\end{figure}

Table \ref{apg_rv_mastertable} contains the RVs we derived for each of the stellar components, while Table \ref{KIC6867766_params} lists the best (least-squares derived) spectroscopic orbital parameters for the 13-day binary.  We rejected some or all of the component RVs measured in five of the epochs due to peak pulling between the three components (generally A and C, but occasionally all three), and these are indicated in Table \ref{apg_rv_mastertable} with ellipses.  Most of our APOGEE spectra have median S/N of $\sim 25-30$ per pixel, but four (2014-04-11, 2014-04-14, 2014-05-11, 2014-06-18) had S/N $\lesssim 10$.  We recovered the M-dwarf signal in each of these low S/N spectra and so retained them in our final orbital solution, albeit with larger RV uncertainties that reflect the poorer data quality.  Fig.\ \ref{kic6867766_sb3} shows the phase-folded RV curves for the AB binary, and residuals from the derived orbital solution.  We also plot the measured RVs for KIC6867766C, which are flat and very close to the systemic velocity of the AB binary.  This suggests that KIC6867766C is a bound companion to the AB binary in a very long-period orbit.  We did not detect significant change in the RV of KIC6867766C, so cannot estimate its orbital period.  We conclude that the KIC6867766 is a likely hierarchical triple system, and an exoplanet false positive corresponding to Scenario 2 in Figure \ref{falseposdiag} (although in this case, the tertiary star is also within the spectrograph's FoV).

\subsection{Survey Efficiency}
\label{qfacsection}
Several high-precision RV instruments are available in the northern hemisphere to confirm \emph{Kepler} exoplanet candidates by achieving RV precisions of a few $\rm{m \; s^{-1}}$, including  HARPS-North, Keck HIRES, SOPHIE, and HET HRS.  These instruments are clearly capable of conducting a reconnaissance survey of \emph{Kepler} KPCs for false positives, but their telescope time is most effectively spent observing robust exoplanet candidates for confirmation and characterization purposes.  Smaller telescopes have been used to measure RVs for hundreds of KOIs through the \emph{Kepler} CFOP (e.g., McDonald 2.7m, Tillinghast 1.5m), and while they have helped identify the best candidates for higher precision RV follow-up, they are limited to a single target at a time: most KOIs have just one or two observations from these facilities.  To compare the efficiency of APOGEE against the larger-aperture telescopes mentioned above, we calculate the integration time per target required to achieve a photon-limited RV precision of $100 ~ \rm{m \; s^{-1}}$.  A quality factor $Q$ is calculated following \citet{bou2001}, adapting instrument parameters summarized in Table \ref{instspecs}.  The quality factor represents the fundamental RV information content of a spectrum, which depends on the number, depths, and widths of spectral features.  The $Q$ values are calculated using BT Settl stellar models for a range of stellar effective temperatures ($T_{\rm{eff}}$), adopting a surface gravity $\log{g} = 5.0$, a solar metallicity, and no rotational broadening.  Including a rotational broadening will affect higher resolution instruments the most, resulting in lower $Q$ values.

From these $Q$ values, it is then possible to calculate the required integration time per object (an ``effective integration time'') to achieve a given, photon-limited RV precision (here taken to be $100 ~ \rm{m \; s^{-1}}$), via:
\begin{equation}
t = \frac{\left( \frac{c}{{{\sigma}}_{\rm{rv}} \; Q \; \sqrt{\pi R^2 \epsilon F}} \right)^2 + t_{\rm{over}}}{n_{\rm{targets}}}
\end{equation}
where $t$ is the effective integration time, $c$ is the speed of light, ${\sigma}_{\rm{rv}}$ is the desired RV precision, $Q$ is the quality factor, $R$ is the telescope's effective aperture radius, $\epsilon$ is the total throughput (as a percentage) of the telescope and instrument, $F$ is the flux in photons per second per unit area, $t_{\rm{over}}$ is the overhead per integration, and $n_{\rm{targets}}$ is the number of targets observed per integration.  Since we are operating in the photon-limited case, we do not include readout noise for the instruments, nor consider sources of systematic uncertainties such as residual moonlight contamination (worse in the optical), or residual telluric lines and sky emission lines (worse in the NIR) -- our interest is in calculating the photon limited case for a direct comparison.

The fluxes are calculated from the BT Settl models after convolving with the appropriate filter transmission function:  Johnson/Bessell $V$\footnote{\url{http://www.ctio.noao.edu/~points/SOIFILTERS/filters/maintext.html}} for HARPS-North, Keck HIRES, SOPHIE, and HET HRS; 2MASS $H$\footnote{\url{http://www.ipac.caltech.edu/2mass/releases/allsky/doc/sec3_1b1.html}} for APOGEE.  The model fluxes are scaled to the zero magnitude level using zero-level fluxes from \citet{bes1998} in $V$ and \citet{coh2003} in 2MASS, which are then further scaled to a desired apparent magnitude.  We calculate the effective integration time for each instrument by including an estimated overhead for detector readout, telescope slew, target acquisition and calibrations.  Since overhead times between integrations depend on a variety of factors, we adopt three minutes as an average overhead time for HARPS-North, Keck HIRES, SOPHIE and HET HRS.  For APOGEE, we use a minimum integration time of 66.7 minutes and an overhead time of 15 minutes, equivalent to the current survey's integration time per visit for each field and its average overhead time between fields.  We do not reduce the integration time for APOGEE below this minimum value because the total integration time is not based on any one target's brightness, but rather an overall integration time required for all targets in a field.

We calculate the required integration times $t$ at a specific $H$ (and corresponding $V$) magnitude for both the single object and multi-object instruments, to allow for more direct comparisons.  We use the median KPC host star flux level of $H \sim 13.5$, along with the appropriately scaled $V$ fluxes for the optical instruments.  The average number of KPC host stars within each APOGEE field that have $ H < 13.5$ is 89, so we use this as the number of KPC hosts that can be observed simultaneously with APOGEE's multiplexing capability for the purposes of comparing against the single-object instruments.  All of these input parameters are summarized in Table \ref{instspecs}.

\begin{figure}
\includegraphics[scale=0.5]{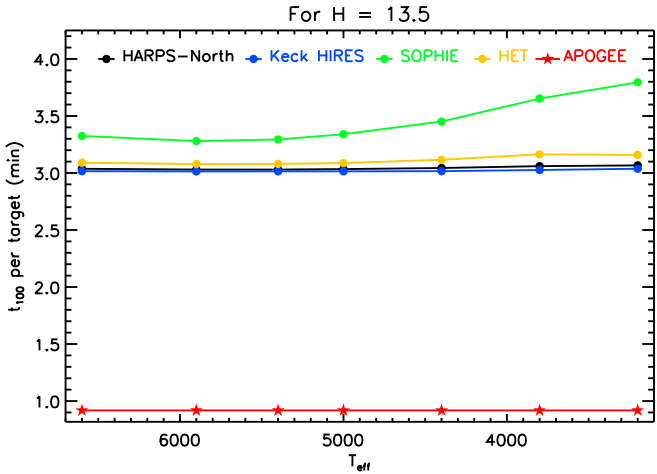}
\caption{Total time (integration + overhead) required to achieve a photon-limited RV precision of $100 ~ \rm{m \; s^{-1}}$ on a per target basis, as a function of stellar $T_{\rm{eff}}$.  Note that at this scale, HARPS-North and Keck HIRES are nearly indistinguishable.  APOGEE's multiplexing capability is the driving factor in reducing the time per target here.\label{exptimesfig}}
\end{figure}

Fig.\ \ref{exptimesfig} displays the required integration times per target to achieve a photon-limited RV precision of $100 ~ \rm{m \; s^{-1}}$ for a variety of spectral types.  At this RV precision, APOGEE is approximately three times as efficient on a per target basis, primarily due to APOGEE's ability to observe multiple KPC hosts simultaneously, and the fact that at this RV precision, the other instruments are dominated by overheads per target.  As the targeted precision level increases, the other instruments become increasingly more efficient compared to APOGEE, reflective of the fact that they are no longer dominated by overheads.  These basic calculations serve to demonstrate how the multiplexing capability of APOGEE enables an efficient survey at modest RV precision compared to the single-object instruments, and that the telescope time for those instruments is best spent on achieving high RV precision to confirm new exoplanets.

\begin{figure}
\includegraphics[scale=0.5]{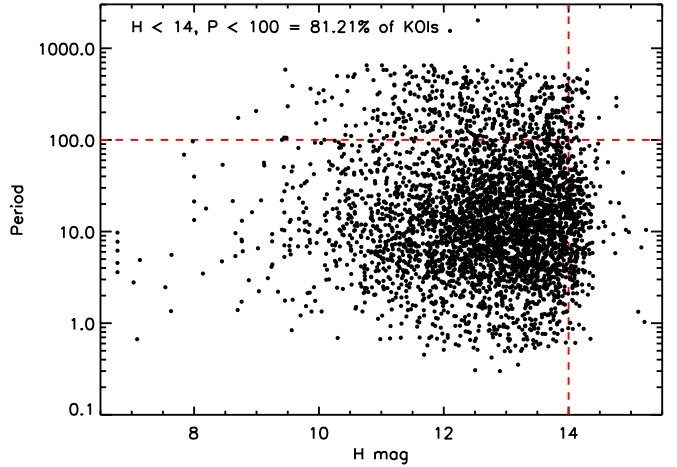}
\caption{Orbital period versus $H$ magnitude for the current KPCs.  The dashed, red lines demarcate $H = 14$ and $P = 100 \; \rm{days}$.  Host stars brighter than this $H$ limit and candidates with periods below 100 days represent $> 80$\% of the current KPC catalog, and are particularly well-suited for APOGEE to characterize the orbits of binaries.  Our SDSS-III and SDSS-IV KOI programs observe targets over more than one year baselines, however, and can detect stellar companions out to the longest KOI orbital periods. \label{hmagvsper}}
\end{figure}

Fig. \ref{hmagvsper} plots the orbital period versus $H$ magnitude for the current KPC catalog.  Host stars with $H < 14$ and $P < 100 \; \rm{days}$ are particularly well-suited for APOGEE to characterize any binary star orbits, and represent more than 80\% of the current KPCs.  Note, however, that even KPCs with longer orbital periods, extending to at least a few hundred days, can be identified as binaries even if the observing baseline does not cover the entire orbital period.  If a more conservative limit of $H < 13$ is applied, more than 47\% of the KPCs would be included.  Scenario 5 in Fig.\ \ref{falseposdiag} relies most heavily on achieving high signal-to-noise ratio observations, and thus has the brightest limiting magnitude within a survey; however, it is also the false positive scenario for which APOGEE is least sensitive.

\section{Discussion and Comparison To Prior Work}
\label{fapsection}
\subsection{Comparison With Other Efforts}
A variety of techniques exist to determine whether a given KPC might be a false positive using only \emph{Kepler} photometric information.  Examination of the transits can be done to ensure that the odd/even transits have the same depth, that there are no ellipsoidal variations, and that the positions of the flux centroids do not vary with brightness changes, all of which are signs that the transit event may be due to a diluted EB \citep{ste2010,tor2011}.  Another method is to generate a large grid of synthetic EB blend scenarios and compare the models to the observed lightcurves via a $\chi^2$ analysis \citep[e.g., BLENDER, ][]{tor2011}, although the technique can be computationally expensive and difficult to apply to all KPCs en masse.  In addition, imaging surveys \citep[e.g., ][]{how2011, ada2012} can be used to inform the photometric analyses described above, particularly to identify fainter stars that exist within the lightcurve aperture, and to search for wide stellar companions to study any relationships between exoplanet properties and host star multiplicity \citep{wan2014}.

Another technique makes use of stellar population synthesis and Galactic models to estimate the probability that a given transit signal is due to an EB.  \citet{mor2011} used such a technique to estimate the false positive probability (FPP) of KPCs, and concluded that 90\% of the KPCs had $\rm{FPPs < 10}$\%, a result that was used by other authors in subsequent statistical analyses of the KPC planet candidates.  However, the technique presented in \citet{mor2011} relied on KPCs to be vetted to the fullest extent possible using the \emph{Kepler} photometry, specifically, the removal of V-shaped transits and searches for faint secondary eclipses, whereas the table of KPCs presented in their paper was not limited to these pre-vetted KPCs.

An updated version of the technique by \citet{mor2012} accounted for more false positive scenarios and clarified the importance of pre-vetting KPCs before performing the statistical analysis.  The updated implementation can be run fairly quickly (of order 10 minutes per star), and has been verified by testing it on confirmed KPCs and known false positive KPCs; however, the technique works best with additional observations (imaging to detect close, visual companions and at least one high resolution spectrum to get coarse stellar parameters).  In addition, the Bayesian modeling is dependent on a variety of model assumptions regarding Galactic structure, stellar population synthesis, distribution of binary properties, and the frequency of exoplanets for various types of stars.  While the framework explicitly accounts for such assumptions through the adopted priors, and is fairly trivial to update when new knowledge is obtained about any of these distributions, direct spectroscopic or photometric observations of false positives are the least model-dependent approach to derive the false positive statistics of KPCs.

The \emph{Kepler} CFOP program has been conducting spectroscopic and imaging campaigns to identify blend sources of KOIs, and, as previously mentioned, have collected thousands of RV measurements for hundreds of KOIs, in addition to high-resolution imaging to search for faint companions unresolved in the \emph{Kepler} photometric aperture.  Our program selected targets independently of those observed by the CFOP program, since our goal was to minimize selection bias in the KOIs we observe.  Our program complements the CFOP in a variety of ways.  The CFOP imaging data can identify wide stellar companions, while our RV measurements over 18-month baselines can detect linear RV trends of intermediate-period companions that are unresolved by AO or lucky imaging.  The abundances of the $\sim$15 elements within APOGEE's $H$ band spectral window can be compared and contrasted with elements accessible in the CFOP optical spectra.

Many of the CFOP spectroscopic observations have obtained just one or two spectra near predicted quadrature phases, where any contaminating spectral lines from a stellar companion are at maximum separation.  This is often sufficient to identify a subset of false positive scenarios: eclipsing stellar companions orbiting the KIC star, for example.  In contrast, each of our KOI targets in SDSS-III/IV (except those few that were part of our \emph{Kepler} EB program) obtain more than twenty APOGEE RVs, which allow us to fully characterize the orbits of any bound companions causing the false positive transit signals.  In the case of eclipsing low-mass secondaries, this further enables a study of the fundamental mass-radius relationship for K and M dwarfs, since precise mass ratios at the 1\% level are achievable.  Our multiple RV measurements can also be used to search for (and place limits on) the presence of any longer-period, non-transiting companions (stellar or otherwise), for the study of multiplicity amongst \emph{Kepler} planet hosts.

Studies have found that the false positive rates for various subsets of KPCs are larger than the ones found from the \citet{mor2011} study, whose quoted statistics are only valid for fully pre-vetted KPCs, and that this rate might differ depending on the orbital period and transit depth of the KPCs.  \citet{col2012} made use of multi-color differential photometry to test for false positives due to diluted EBs whose components have sufficiently different colors.  They observed a total of four KPCs that had short periods ($P < 6 \; \rm{days}$) and small radii ($R_{p} < 5 \; R_{\oplus}$), and found evidence that two of the four were likely due to diluted EBs, excluding an overall false positive rate of 10\% with 99\% confidence.  \citet{san2012} collected spectroscopic RVs of 33 giant planet KPCs using the SOPHIE spectrograph and found a false positive probability of at least $35$\% within their sample, where a majority of false positives were due to EBs.  Their sample size of 33 KPCs was partly limited by their telescope resource:  a single-object spectrograph observing in the optical using a 2m-class telescope, corresponding to an effective magnitude limit in the \emph{Kepler} bandpass of $K_p \lesssim 14.7$.  This magnitude limit removes almost half of the total KPCs.  Utilizing a multi-object, NIR spectrograph, such as APOGEE, increases the rate of data collection while also increasing the total number of KPCs able to be observed.

\subsection{Abundances of KOI Host Stars}
Beyond identifying false positive KPCs as binaries, a variety of additional science projects can be done with the NIR APOGEE spectra.  One such example is the study of chemical abundance patterns in planet host stars compared to stars not known to host planets.  One of APOGEE's primary goals is to measure the chemical abundances of many elements to study stellar populations within the Milky Way.  These abundances are measured using the APOGEE Stellar Parameters and Chemical Abundances Pipeline \citep[ASPCAP,][]{gar2015}, which consists of a suite of software codes to analyze, in an automated fashion, the APOGEE spectra.  The main component of the code is FERRE\footnote{FERRE is available at \url{http://hebe.as.utexas.edu/ferre}}, a Fortran optimization code that searches for the set of parameters that best match each APOGEE spectrum.  FERRE was originally developed in the context of low-resolution SDSS spectroscopy \citep{all2004, all2006}, and has subsequently evolved and been used in other contexts \citep{all2008, all2009, bro2012, kil2012}.  The APOGEE band (1.5-1.7 $\mu$m) is rich in transitions from many elements in cool stars. Abundances for 15 elements can be derived from sufficiently high resolution ($R>20000$) and signal-to-noise per pixel (S/N$>100$) spectra in this spectral window:  C, N, O, Na, Mg, Al, Si, K, Ca, Ti, V, Mn, Fe, Ni, Cr.  A S/N level approaching 100 per pixel is expected to be achieved out to $H \sim 14$ by coadding the multiple visits.

\begin{figure}
\includegraphics[scale=0.5]{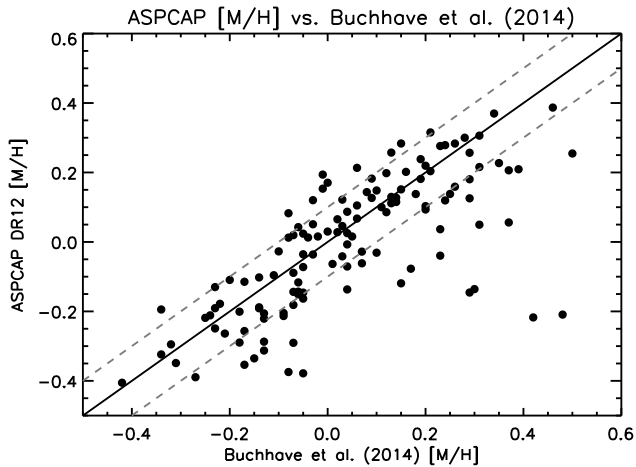}
\caption{Comparison of [M/H] values from ASPCAP DR12 and \citet{buc2014} for stars in common.  The ASPCAP values are taken from the `''uncalibrated'' (initial fit) values (``FPARAM'' array).  We find that a majority of these targets fall within 0.1 dex of the \citet{buc2014} values (grey dashed lines).  These boundaries also represent the typical 1$\sigma$ uncertainties for both sets of values.  There are not enough targets with ``externally calibrated'' values (``PARAM'' array) in ASPCAP DR12 to make a statement about how those compare to \citet{buc2014}. \label{buccomp}}
\end{figure}

A detailed analysis of elemental abundances for KOIs with and without exoplanets is beyond the scope of this introductory paper.  To get some sense of the metallicites coming from the automated ASPCAP pipeline, we have compared the ``uncalibrated'' [M/H] values from ASPCAP in DR12 with those found in \citet{buc2014} for stars in common.  The ``uncalibrated'' values are the parameters that come from the initial fit.  A subset of targets have additional, external calibrations applied to their stellar parameters (such as metallicities of clusters from the literature).  We refer the reader to the ASPCAP DR12 documentation\footnote{\url{http://www.sdss.org/dr12/irspec/parameters/}} for full details.  There are not enough targets that have had external calibrations applied to make any statement regarding their agreement.  However, we find a total of 128 KOI host stars observed in common that have ``uncalibrated'' [M/H] values in DR12.  We find the agreement to be promising (Fig.\ \ref{buccomp}): a majority of targets agree within 0.1 dex (grey lines in Fig.\ \ref{buccomp}), despite the fact that ASPCAP has been calibrated primarily to work on bright giants.  We have not made any cuts based on ASPCAP processing flags; this is a comparison using DR12 [M/H] values ``as they are'' versus the \citet{buc2014} values.  As such, the relation in Fig.\ \ref{buccomp} should be considered as preliminary, and likely to be improved upon in future analyses.

There are tantalizing hints for different heavy element patterns in planet hosts relative to the field that could be induced by preferential removal of heavy elements in the disk.  It's been shown that stars hosting Jovian-mass planets tend to be more metal-rich than stars with only Neptunian-mass planets \citep{sou2008,ghe2010,sou2011}.  The overall shift in values of [Fe/H] between stars with Jovian- versus Neptunian-mass planets is 0.20 dex, which is significant and indicates that the metallicity populations for stars with Jovian-mass planets are not the same as those which host the smaller Neptunian-mass planets.  Others have used the \emph{Kepler} sample to extend the planet-metallicity correlation down to terrestrial-sized planets \citep{buc2012}, and have found that terrestrial-sized planets fall into well-defined host-star metallicity regimes \citep{buc2014}.  These results suggest that metallicity may also influence the distribution of planetary masses within extrasolar systems.  In addition to stellar metallicity, there are also suggestions that stellar mass plays a role, such that the dominant planetary mass decreases as the parent star's mass decreases \citep{ghe2010,may2011}, at least for main sequence stars.  While massive subgiant and giant stars show trends similar to main sequence stars, low-mass giants show a different behavior \citep{mal2013}.  It is likely that some combination of stellar mass and metallicity influences the type of planetary system that will form \citep{joh2010}.  APOGEE can provide a statistically significant control sample for such studies.

In addition to overall metallicity and stellar mass playing a role, the detailed chemistry of the parent cloud in which the system forms may also hold clues to further understanding planet formation.  Recent findings suggest that specific abundance patterns, such as Mg/Fe, may influence the likelihood that a star hosts an underlying planetary system \citep{adi2012a}, and that enhancement in alpha elements may favor the formation of rocky planets, even for stars with low iron abundances \citep{adi2012b}.  The C/O ratio in the parent cloud is also found in some studies to be enriched in planet hosting systems, with C/O ratios $> 0.8$ \citep{del2010,pet2011}, however, these results have been questioned by \citet{for2012}, and have not been confirmed by other groups \citep{nis2013,tes2014}. \citet{bru2011} find that the silicon abundance (and not the oxygen abundance) is a key element, as they find that their planet detection rate depends strongly on the silicon abundance of the host star.  A difference in the Si abundance is also found for the XO-2 binary host stars, where XO-2N is found to be enhanced relative to XO-2S \citep{tes2015}.  \citet{mel2009}, \citet{gon2010}, and \citet{sch2011} present intriguing results suggesting that low-amplitude chemical signatures point to selective accretion or depletion of refractory and volatile elements in stellar exoplanetary hosts.  In particular, trends of abundances with condensation temperature ($T_{\rm{cond}}$) are used as diagnostics, and these can be defined from the abundances of the 15 chemical elements covered by APOGEE, which include C, N, O (volatiles) and Si, Ti and Al (refractories).  The investigation of such trends in samples of \emph{Kepler} stars with confirmed planets of different masses, and including the smallest planets to date, provides an unprecedented database in order to probe the importance of $T_{\rm{cond}}$ trends in this context.

\section{Summary}
\label{summarysection}
In this paper, we highlight the importance of an RV survey of KPCs to better determine the false positive rate, and demonstrate that APOGEE can efficiently conduct such a survey of KPCs to identify, and in many cases characterize the orbits of, false positive KPCs.  We have shown that the APOGEE instrument is capable of achieving an RV precision of $\sim 100 ~ \rm{m \; s^{-1}}$ using observations of the confirmed exoplanet host star KIC 6448890, as well as the \emph{Kepler} EB KIC 01571511, which produces planet-sized eclipses and has an $H$ magnitude similar to many KPCs.  We find that the transit signal of KIC 3848972 is not caused by a blue, stellar secondary orbiting the primary star, and do not find any evidence of a faint, red companion in Keck AO images that could be the source of the \emph{Kepler} transit events.  Further investigation is merited before the true nature of this KOI can be confidently identified.  We find HET RV variations that phase to the KOI period and ephemeris for KIC 3861595, but our APOGEE RVs are inconsistent with the HET RV semiamplitude, and we find evidence of line bisector variations.  This target was part of our EB program in SDSS, so we only have a few spectra to work with, and the RV uncertainties have not been fully vetted for rapdily rotating stars such as this one.  As such, we can not definitively determine the nature of this KPC, and urge more spectra be obtained to examine both the RV and line bisector variations.  Finally, we find that KIC 6867766 is a triple system, composed of an F+M EB and a wide, bound G dwarf tertiary.  The F+M EB phases to the KOI period and ephemeris, and the diluted eclipses are the source of the KPC transits.  As such, we can confidently identify this KOI as a false positive exoplanet candidate.

Not only can the data from such a survey be used to determine the false positive rate of KPCs and vet the sample to identify the best candidates for high-precision RV observations, but it will enable ancillary science projects in fundamental stellar astrophysics though observations of EBs, studies of intrinsically rare short-period companions (such as brown dwarfs), and detailed chemical abundances of exoplanet host stars.  At the precision level of $100 ~ \rm{m \; s^{-1}}$, APOGEE is a more efficient instrument compared to HARPS-North, Keck HIRES, and SOPHIE, due to its multiplexing capability and because the single-object spectrographs are dominated by overheads.  Our survey to detect false positives refines the target selection for higher precision RV instruments, enabling them to focus on the best exoplanet candidates.  It will allow for improved statistical studies of the \emph{Kepler} exoplanet population by determining the false positive rate of KPCs due to physically-bound binaries, as well as any trends in the false positive rate with orbital period or stellar properties.

\acknowledgements
We thank the anonymous referee for detailed comments that improved the quality of this publication, especially their suggestion to conduct more analysis on the KOI-4 RVs.  We acknowledge support from NSF award AST1006676, AST 1126413, and AST 1310885.  This work was partially supported by funding from the Center for Exoplanets and Habitable Worlds. The Center for Exoplanets and Habitable Worlds is supported by the Pennsylvania State University, the Eberly College of Science, and the Pennsylvania Space Grant Consortium.  CAP acknowledges funding from the Spanish Ministry of Economy and Competitiveness (MINECO) through grant AYA2011-26244.  This research has made use of the SIMBAD database, operated at CDS, Strasbourg, France.  This publication makes use of data products from the Two Micron All Sky Survey, which is a joint project of the University of Massachusetts and the Infrared Processing and Analysis Center/California Institute of Technology, funded by the National Aeronautics and Space Administration and the National Science Foundation.  Some of the data presented in this paper were obtained from the Mikulski Archive for Space Telescopes (MAST). STScI is operated by the Association of Universities for Research in Astronomy, Inc., under NASA contract NAS5-26555. Support for MAST for non-HST data is provided by the NASA Office of Space Science via grant NNX13AC07G and by other grants and contracts.  This paper includes data collected by the \emph{Kepler} mission. Funding for the \emph{Kepler} mission is provided by the NASA Science Mission directorate.  This research has made use of the NASA Exoplanet Archive, which is operated by the California Institute of Technology, under contract with the National Aeronautics and Space Administration under the Exoplanet Exploration Program.

This work was based on observations with the SDSS 2.5-meter telescope.  Funding for SDSS-III has been provided by the Alfred P. Sloan Foundation, the Participating Institutions, the National Science Foundation, and the U.S. Department of Energy Office of Science. The SDSS-III web site is \url{http://www.sdss3.org/}.  SDSS-III is managed by the Astrophysical Research Consortium for the Participating Institutions of the SDSS-III Collaboration including the University of Arizona, the Brazilian Participation Group, Brookhaven National Laboratory, University of Cambridge, Carnegie Mellon University, University of Florida, the French Participation Group, the German Participation Group, Harvard University, the Instituto de Astrofisica de Canarias, the Michigan State/Notre Dame/JINA Participation Group, Johns Hopkins University, Lawrence Berkeley National Laboratory, Max Planck Institute for Astrophysics, Max Planck Institute for Extraterrestrial Physics, New Mexico State University, New York University, Ohio State University, Pennsylvania State University, University of Portsmouth, Princeton University, the Spanish Participation Group, University of Tokyo, University of Utah, Vanderbilt University, University of Virginia, University of Washington, and Yale University.

\appendix
\section{TRILEGAL Input Parameters}
\label{appendix1}
The parameters used to calculate the background EB blend probability of KIC 3861595 are summarized in Table \ref{trilegal_params}.

%%%%%%%%%%%%%%%%%% tables here %%%%%%%%%%%%%%%%%%

\clearpage

\begin{deluxetable}{lllllllll}
\tabletypesize{\scriptsize}
\tablecaption{RVs for KIC Stars - All RVs in $\rm{km \; s^{-1}}$\label{apg_rv_mastertable}}
\tablewidth{0pt}
\tablehead{
\colhead{KIC ID} & \colhead{BJD\_TDB} & \colhead{$\textrm{RV}_A$} & \colhead{$1\sigma$} & \colhead{$\textrm{RV}_B$} & \colhead{$1\sigma$} & \colhead{$\textrm{RV}_C$} & \colhead{$1\sigma$} & \colhead{Instrument}
}
\startdata
1571511 & 2455811.61304 &  -24.401 &    0.153 & $\cdots$ & $\cdots$ & $\cdots$ & $\cdots$ & APOGEE \\
1571511 & 2455840.59327 &  -26.348 &    0.115 & $\cdots$ & $\cdots$ & $\cdots$ & $\cdots$ & APOGEE \\
1571511 & 2455851.57845 &  -18.927 &    0.105 & $\cdots$ & $\cdots$ & $\cdots$ & $\cdots$ & APOGEE \\
\hline
3848972 & 2455811.61297 &  -19.943 &    0.157 & $\cdots$ & $\cdots$ & $\cdots$ & $\cdots$ & APOGEE \\
3848972 & 2455840.59327 &  -20.161 &    0.153 & $\cdots$ & $\cdots$ & $\cdots$ & $\cdots$ & APOGEE \\
3848972 & 2455851.57848 &  -19.641 &    0.150 & $\cdots$ & $\cdots$ & $\cdots$ & $\cdots$ & APOGEE \\
\hline
3861595 & 2455789.84195 &  -23.052 &    0.091 & $\cdots$ & $\cdots$ & $\cdots$ & $\cdots$ & HET    \\
3861595 & 2455796.82710 &  -23.827 &    0.075 & $\cdots$ & $\cdots$ & $\cdots$ & $\cdots$ & HET    \\
3861595 & 2455797.80608 &  -23.204 &    0.081 & $\cdots$ & $\cdots$ & $\cdots$ & $\cdots$ & HET    \\
3861595 & 2455801.80843 &  -23.194 &    0.098 & $\cdots$ & $\cdots$ & $\cdots$ & $\cdots$ & HET    \\
3861595 & 2455803.80347 &  -24.285 &    0.103 & $\cdots$ & $\cdots$ & $\cdots$ & $\cdots$ & HET    \\
3861595 & 2455813.70317 &  -22.543 &    0.566 & $\cdots$ & $\cdots$ & $\cdots$ & $\cdots$ & APOGEE \\
3861595 & 2455823.72718 &  -23.628 &    0.469 & $\cdots$ & $\cdots$ & $\cdots$ & $\cdots$ & APOGEE \\
3861595 & 2455840.66180 &  -22.930 &    0.541 & $\cdots$ & $\cdots$ & $\cdots$ & $\cdots$ & APOGEE \\
3861595 & 2455849.57900 &  -24.600 &    0.434 & $\cdots$ & $\cdots$ & $\cdots$ & $\cdots$ & APOGEE \\
3861595 & 2455851.64939 &  -21.337 &    0.433 & $\cdots$ & $\cdots$ & $\cdots$ & $\cdots$ & APOGEE \\
3861595 & 2455866.56998 &  -21.608 &    0.461 & $\cdots$ & $\cdots$ & $\cdots$ & $\cdots$ & APOGEE \\
\hline
6448890 & 2456368.99828 &  -55.391 &    0.117 & $\cdots$ & $\cdots$ & $\cdots$ & $\cdots$ & APOGEE \\
6448890 & 2456411.92027 &  -55.513 &    0.105 & $\cdots$ & $\cdots$ & $\cdots$ & $\cdots$ & APOGEE \\
6448890 & 2456557.73343 &  -55.598 &    0.103 & $\cdots$ & $\cdots$ & $\cdots$ & $\cdots$ & APOGEE \\
6448890 & 2456559.72336 &  -55.644 &    0.106 & $\cdots$ & $\cdots$ & $\cdots$ & $\cdots$ & APOGEE \\
6448890 & 2456560.72108 &  -55.571 &    0.104 & $\cdots$ & $\cdots$ & $\cdots$ & $\cdots$ & APOGEE \\
6448890 & 2456584.63225 &  -55.582 &    0.104 & $\cdots$ & $\cdots$ & $\cdots$ & $\cdots$ & APOGEE \\
6448890 & 2456585.63076 &  -55.644 &    0.105 & $\cdots$ & $\cdots$ & $\cdots$ & $\cdots$ & APOGEE \\
6448890 & 2456757.89294 &  -55.622 &    0.107 & $\cdots$ & $\cdots$ & $\cdots$ & $\cdots$ & APOGEE \\
6448890 & 2456758.90229 &  -55.801 &    0.142 & $\cdots$ & $\cdots$ & $\cdots$ & $\cdots$ & APOGEE \\
6448890 & 2456760.90571 &  -55.586 &    0.112 & $\cdots$ & $\cdots$ & $\cdots$ & $\cdots$ & APOGEE \\
6448890 & 2456761.87281 &  -55.573 &    0.139 & $\cdots$ & $\cdots$ & $\cdots$ & $\cdots$ & APOGEE \\
6448890 & 2456762.86860 &  -55.621 &    0.111 & $\cdots$ & $\cdots$ & $\cdots$ & $\cdots$ & APOGEE \\
6448890 & 2456763.88112 &  -55.581 &    0.109 & $\cdots$ & $\cdots$ & $\cdots$ & $\cdots$ & APOGEE \\
6448890 & 2456783.83567 &  -55.712 &    0.112 & $\cdots$ & $\cdots$ & $\cdots$ & $\cdots$ & APOGEE \\
6448890 & 2456784.82195 &  -55.781 &    0.133 & $\cdots$ & $\cdots$ & $\cdots$ & $\cdots$ & APOGEE \\
6448890 & 2456785.82543 &  -55.702 &    0.108 & $\cdots$ & $\cdots$ & $\cdots$ & $\cdots$ & APOGEE \\
6448890 & 2456786.79845 &  -55.590 &    0.113 & $\cdots$ & $\cdots$ & $\cdots$ & $\cdots$ & APOGEE \\
6448890 & 2456787.80934 &  -55.640 &    0.107 & $\cdots$ & $\cdots$ & $\cdots$ & $\cdots$ & APOGEE \\
6448890 & 2456788.84307 &  -55.679 &    0.118 & $\cdots$ & $\cdots$ & $\cdots$ & $\cdots$ & APOGEE \\
6448890 & 2456812.74509 &  -55.620 &    0.111 & $\cdots$ & $\cdots$ & $\cdots$ & $\cdots$ & APOGEE \\
6448890 & 2456814.75547 &  -55.615 &    0.114 & $\cdots$ & $\cdots$ & $\cdots$ & $\cdots$ & APOGEE \\
6448890 & 2456815.78552 &  -55.607 &    0.107 & $\cdots$ & $\cdots$ & $\cdots$ & $\cdots$ & APOGEE \\
6448890 & 2456816.76627 &  -55.710 &    0.119 & $\cdots$ & $\cdots$ & $\cdots$ & $\cdots$ & APOGEE \\
6448890 & 2456817.76198 &  -55.632 &    0.109 & $\cdots$ & $\cdots$ & $\cdots$ & $\cdots$ & APOGEE \\
6448890 & 2456818.76458 &  -55.609 &    0.110 & $\cdots$ & $\cdots$ & $\cdots$ & $\cdots$ & APOGEE \\
6448890 & 2456819.76222 &  -55.567 &    0.109 & $\cdots$ & $\cdots$ & $\cdots$ & $\cdots$ & APOGEE \\
6448890 & 2456820.75601 &  -55.666 &    0.108 & $\cdots$ & $\cdots$ & $\cdots$ & $\cdots$ & APOGEE \\
\hline
6867766 & 2456557.73337 & $\cdots$ & $\cdots$ & $\cdots$ & $\cdots$ & $\cdots$ & $\cdots$ & APOGEE \\
6867766 & 2456559.72331 &   38.598 &    0.383 &  -60.127 &    2.882 &    6.928 &    0.664 & APOGEE \\
6867766 & 2456560.72103 &   40.497 &    0.365 &  -65.538 &    6.014 &    8.042 &    0.653 & APOGEE \\
6867766 & 2456584.63222 &   28.939 &    0.399 &  -36.948 &    3.811 &    7.788 &    0.780 & APOGEE \\
6867766 & 2456585.63072 &   37.993 &    0.446 &  -64.834 &    4.138 &    8.956 &    0.754 & APOGEE \\
6867766 & 2456757.89298 & $\cdots$ & $\cdots$ & $\cdots$ & $\cdots$ & $\cdots$ & $\cdots$ & APOGEE \\
6867766 & 2456758.90233 &    1.277 &    1.679 &   46.983 &    9.919 &    8.567 &    3.754 & APOGEE \\
6867766 & 2456760.90575 &  -20.442 &    0.482 &   88.944 &    3.869 &   10.516 &    0.778 & APOGEE \\
6867766 & 2456761.87284 &  -17.315 &    1.389 &   79.738 &   11.967 &   10.500 &    3.354 & APOGEE \\
6867766 & 2456762.86864 &  -10.374 &    0.432 &   66.784 &    3.542 &    9.447 &    0.760 & APOGEE \\
6867766 & 2456763.88116 & $\cdots$ & $\cdots$ &   28.563 &    3.463 & $\cdots$ & $\cdots$ & APOGEE \\
6867766 & 2456783.83568 &   15.664 &    0.590 &   -3.902 &    3.310 &   11.547 &    1.018 & APOGEE \\
6867766 & 2456784.82197 & $\cdots$ & $\cdots$ &   37.085 &    3.697 & $\cdots$ & $\cdots$ & APOGEE \\
6867766 & 2456785.82544 &  -12.960 &    0.422 &   73.438 &    3.067 &    9.843 &    0.759 & APOGEE \\
6867766 & 2456786.79846 &  -21.181 &    0.542 &   89.949 &    5.130 &    9.443 &    0.887 & APOGEE \\
6867766 & 2456787.80935 &  -20.018 &    0.398 &   92.288 &    4.554 &    9.495 &    0.656 & APOGEE \\
6867766 & 2456788.84309 &  -12.259 &    1.175 &   55.283 &    9.284 &    7.295 &    1.769 & APOGEE \\
6867766 & 2456812.74507 &  -20.349 &    0.501 &   90.837 &    6.466 &    8.471 &    0.870 & APOGEE \\
6867766 & 2456814.75546 &  -10.567 &    0.455 &   70.185 &    6.978 &    9.955 &    0.891 & APOGEE \\
6867766 & 2456815.78551 & $\cdots$ & $\cdots$ &   32.014 &    3.443 & $\cdots$ & $\cdots$ & APOGEE \\
6867766 & 2456816.76626 &   16.157 &    0.886 &   -7.116 &    4.691 &    9.461 &    1.656 & APOGEE \\
6867766 & 2456817.76197 &   27.340 &    0.445 &  -35.030 &    4.168 &   11.971 &    1.059 & APOGEE \\
6867766 & 2456818.76456 &   37.006 &    0.398 &  -59.225 &    3.500 &    8.936 &    0.702 & APOGEE \\
6867766 & 2456819.76220 &   40.742 &    0.422 &  -65.495 &    3.783 &    9.396 &    0.780 & APOGEE \\
6867766 & 2456820.75599 &   38.303 &    0.356 &  -57.968 &    3.165 &    8.577 &    0.668 & APOGEE \\
\enddata
\end{deluxetable}

\clearpage

\begin{deluxetable}{lll}
\tabletypesize{\scriptsize}
\tablecaption{KIC 6867766 A+B Orbital Parameters\label{KIC6867766_params}}
\tablewidth{0pt}
\tablehead{
\colhead{Parameter} & \colhead{Value} & \colhead{$1\sigma$}
}
\startdata
$P$ (days) & 12.964712 & (fixed at \emph{Kepler} value) \\
$T_p$ & 2456746.58 & 0.21 \\
$e$ & 0.0553 & 0.0054 \\
$\omega$ (deg) & 128.6 & 5.7 \\
$K_A$ ($\rm{km \; s^{-1}}$) & 30.77 & 0.14 \\
$K_B$ ($\rm{km \; s^{-1}}$) & 79.4 & 1.2 \\
$\gamma$ ($\rm{km \; s^{-1}}$) & 10.87 & 0.14 \\
$M_B / M_A$ & 0.3877 & 0.0060 \\ 
\enddata
\end{deluxetable}

%\clearpage

\begin{deluxetable}{lrrrrrrr}
\tabletypesize{\scriptsize}
\tablecaption{Instrument Parameters For $Q$ Factor Calculation\label{instspecs}}
\tablewidth{0pt}
\tablehead{
\colhead{Instrument} & \colhead{Resolution} & \colhead{${\lambda}_{\rm{min}}$} & \colhead{${\lambda}_{\rm{max}}$} & \colhead{Eff. Aperture} & \colhead{Total Throughput} & \colhead{Overhead} & \colhead{\# Targets / Obs.} \\
\colhead{~} & \colhead{~} & \colhead{$\left( \rm{\AA} \right)$} & \colhead{$\left( \rm{\AA} \right)$} & \colhead{Radius $\left( \rm{m} \right)$} & \colhead{$\left( {\%} \right)$} & \colhead{$\left( \rm{min} \right)$} & \colhead{$\left( H \sim 13.5 \right)$}
}
\startdata
HARPS-North\tablenotemark{a} & 115000 & 3830 & 6930 & 3.58 & 8 & 3 & 1\\
Keck HIRES\tablenotemark{b} & 55000 & 5000 & 6200 & 10.0 & 13 & 3 & 1\\
SOPHIE\tablenotemark{c} & 75000 & 3820 & 6930 & 1.93 & 4 & 3 & 1\\
HET\tablenotemark{d} & 30000 & 4076 & 7838 & 9.2 & 3 & 3 & 1\\
APOGEE\tablenotemark{e} & 22500 & 15100 & 17000 & 2.12 & 16 & 15 & 89\\
\enddata
\tablenotetext{a}{Instrument parameters taken from \url{http://www.tng.iac.es/instruments/harps/}.}
\tablenotetext{b}{Instrument specs as reported in \citet{joh2011}.  Efficiency is taken to be 18\% from \url{http://www2.keck.hawaii.edu/inst/hires/throughput.pdf}, minus an additional 30\% loss due to absorption from the iodine cell.}
\tablenotetext{c}{Instrument parameters taken from \url{http://www.obs-hp.fr/guide/sophie/sophie-info.html}.}
\tablenotetext{d}{Instrument parameters are for the 316g5936 cross-disperser in $R = 30000$ mode using ThAr as a wavelength calibration.  Although most planet work has been done using the $R = 60000$ mode, this resolution is not required to achieve $100 \; \rm{m \; s^{-1}}$ RV precision.  Efficiency taken from the HRS exposure time calculator \url{http://het.as.utexas.edu/HET/hetweb/Instruments/HRS/exp/exp_calc.html}, and is calculated at the center of the telescope's observability track.}
\tablenotetext{e}{Instrument parameters taken from \citet{wil2012}.  Effective aperture radius includes a 30\% loss due to obstruction of the 2.5m diameter \citep{gun2006}.}
\end{deluxetable}

\clearpage

\begin{deluxetable}{ll}
\tabletypesize{\scriptsize}
\tablecaption{TRILEGAL Parameters Used In KIC 3861595 Background EB Blend Probability Calculation \label{trilegal_params}}
\tablewidth{0pt}
\tablehead{
\colhead{Parameter} & \colhead{Value}
}
\startdata
Distance modulus resolution of Galaxy components & 0.1 mag \\
IMF for single stars & Chabrier lognormal \\
Binary fraction & 0.3 \\
Binary mass ratios & 0.7 to 0.1 \\
Extinction model & Exponential disk of form $\exp{\left(-\left|z\right|/h_{z,\rm{dust}}\right)} * \exp{\left(-R/h_{R,\rm{dust}}\right)}$ \\
Extinction model $h_{z,\rm{dust}}$ & 110 pc \\
Extinction model $h_{R,\rm{dust}}$ & 100000 pc \\
Extinction calibration at infinity & 0.0378 \\
$1\sigma$ extinction dispersion & 0. \\
Solar Galactocentric radius $R_{\odot}$ & 8700 pc \\
Solar height above the disk $z_{\odot}$ & 24.2 pc \\
Thin disk model & squared hyperbolic secant \\
Thin disk $z_0$ & 94.6902 pc \\
Thin disk $t_0$ & 5.55079E9 yr \\
Thin disk $\alpha$ & 1.6666 \\
Thin disk $h_{R,d}$ & 2913.36 pc \\
Thin disk radial cutoffs & 0, 15000 pc \\
Thin disk $\Sigma_d\left(\odot\right)$ & 55.4082 $M_{\odot} \; pc^{-2}$ \\
Thin disk SFR+AMR & 2-step SFR + Fuhrman's AMR + $\alpha$ enhancement with age(yr) = $0.735097t + 0$ \\
Thick disk model & squared hyperbolic secant \\
Thick disk $h_{z,td}$ & 800 pc \\
Thick disk $h_{R,td}$ & 2394.07 pc \\
Thick disk radial cutoffs & 0, 15000 pc \\
Thick disk $\Omega_{td}\left(\odot\right)$ & 0.001 $M_{\odot} \; pc^{-3}$ \\
Thick disk SFR+AMR & 11-12 Gyr const. SFR + Z=0.008 with $\sigma$[M/H] = 0.1 dex with age(yr) = $t + 0$ \\
Halo model & Oblate $r^{1/4}$ spheroid \\
Halo $r_h$ & 2698.93 pc \\
Halo $q_h$ & 0.583063 \\
Halo $\Omega_h\left(\odot\right)$ & 0.000100397 $M_{\odot} \; pc^{-3}$ \\
Halo SFR+AMR & 12-13 Gyr + Ryan \& Norris [M/H] distribution with age (yr) = $t + 0$ \\
Bulge model & triaxial bulge \\
Bulge $a_m$ & 2500 pc \\
Bulge $a_0$ & 95 pc \\
Bulge $y/x$ axial ratio $\eta$ & 0.68 \\
Bulge $z/x$ axial ratio $\xi$ & 0.31 \\
Bulge Sun-GC-bar angle $\phi_0$ & $15^\circ$ \\
Bulge $\Omega_b\left(\rm{GC}\right)$ & 406 $M_{\odot} \; pc^{-3}$ \\
Bulge SFR+AMR & 10 Gyr, Zoccali et al. 2003 [M/H] + 0.3 dex with age(yr) = $t - \rm{2E9}$ \\
\enddata
\end{deluxetable}

%%%%%%%%%%%%%%%%%% end tables %%%%%%%%%%%%%%%%%%

%%%%%%%%%%%%%%%%%% figures here %%%%%%%%%%%%%%%%%%

\clearpage

%%%%%%%%%%%%%%%%%% end figures %%%%%%%%%%%%%%%%%%


\begin{thebibliography}{}
\bibitem[Adams et al. (2012)]{ada2012} Adams, E.~R., Ciardi, D.~R., Dupree, A.~K., et al.\ 2012, \aj, 144, 42
\bibitem[Adibekyan et al. (2012a)]{adi2012a} Adibekyan, V.~Z., Santos, N.~C., Sousa, S.~G., et al.\ 2012, \aap, 543, A89
\bibitem[Adibekyan et al. (2012b)]{adi2012b} Adibekyan, V.~Z., Delgado Mena, E., Sousa, S.~G., et al.\ 2012, arXiv:1209.6272
\bibitem[Allard et al. (2011)]{ala2011} Allard, F., Homeier, D., \& Freytag, B.\ 2011, 16th Cambridge Workshop on Cool Stars, Stellar Systems, and the Sun, 448, 91
\bibitem[Allende Prieto (2004)]{all2004} Allende Prieto, C.\ 2004, Astronomische Nachrichten, 325, 604
\bibitem[Allende Prieto et al. (2006)]{all2006} Allende Prieto, C., Beers, T.~C., Wilhelm, R., et al.\ 2006, \apj, 636, 804
\bibitem[Allende Prieto et al. (2008a)]{all2008} Allende Prieto, C., Sivarani, T., Beers, T.~C., et al.\ 2008, \aj, 136, 2070
\bibitem[Allende Prieto et al. (2008b)]{all2008b} Allende Prieto, C., Majewski, S.~R., Schiavon, R., et al.\ 2008, Astronomische Nachrichten, 329, 1018
\bibitem[Allende Prieto et al. (2009)]{all2009} Allende Prieto, C., Hubeny, I., \& Smith, J.~A.\ 2009, \mnras, 396, 759
\bibitem[Batalha et al. (2013)]{bat2013} Batalha, N.~M., Rowe, J.~F., Bryson, S.~T., et al.\ 2013, \apjs, 204, 24
\bibitem[Bender \& Simon (2008)]{ben2008} Bender, C.~F., \& Simon, M.\ 2008, \apj, 689, 416 
\bibitem[Bender et al. (2012)]{ben2012} Bender, C.~F., Mahadevan, S., Deshpande, R., et al.\ 2012, \apjl, 751, L31
\bibitem[Bessell et al. (1998)]{bes1998} Bessell, M.~S., Castelli, F., \& Plez, B.\ 1998, \aap, 333, 231
\bibitem[Borucki et al. (2010)]{bor2010} Borucki, W.~J., Koch, D., Basri, G., et al.\ 2010, Science, 327, 977
\bibitem[Borucki et al. (2011a)]{bor2011a} Borucki, W.~J., Koch, D.~G., Basri, G., et al.\ 2011, \apj, 728, 117 
\bibitem[Borucki et al. (2011b)]{bor2011b} Borucki, W.~J., Koch, D.~G., Basri, G., et al.\ 2011, \apj, 736, 19
\bibitem[Bouchy et al. (2001)]{bou2001} Bouchy, F., Pepe, F., \& Queloz, D.\ 2001, \aap, 374, 733
\bibitem[Brown et al. (2011)]{bro2011} Brown, T.~M., Latham, D.~W., Everett, M.~E., \& Esquerdo, G.~A.\ 2011, \aj, 142, 112
\bibitem[Brown et al. (2012)]{bro2012} Brown, W.~R., Kilic, M., Allende Prieto, C., \& Kenyon, S.~J.\ 2012, \apj, 744, 142
\bibitem[Brugamyer et al. (2011)]{bru2011} Brugamyer, E., Dodson-Robinson, S.~E., Cochran, W.~D., \& Sneden, C.\ 2011, \apj, 738, 97
\bibitem[Buchhave et al. (2012)]{buc2012} Buchhave, L.~A., Latham, D.~W., Johansen, A., et al.\ 2012, \nat, 486, 375
\bibitem[Buchhave et al. (2014)]{buc2014} Buchhave, L.~A., Bizzarro, M., Latham, D.~W., et al.\ 2014, \nat, 509, 593 
\bibitem[Butler et al. (1996)]{but1996} Butler, R.~P., Marcy, G.~W., Williams, E., et al.\ 1996, \pasp, 108, 500
\bibitem[Caldwell et al. (2010)]{cal2010} Caldwell, D.~A., Kolodziejczak, J.~J., Van Cleve, J.~E., et al.\ 2010, \apjl, 713, L92
\bibitem[Carter et al. (2011)]{car2011} Carter, J.~A., Fabrycky, D.~C., Ragozzine, D., et al.\ 2011, Science, 331, 562
\bibitem[Chaplin et al. (2011)]{cha2011} Chaplin, W.~J., Kjeldsen, H., Christensen-Dalsgaard, J., et al.\ 2011, Science, 332, 213
\bibitem[Christiansen et al. (2012)]{chr2012} Christiansen, J.~L., Jenkins, J.~M., Barclay, T.~S., et al.\ 2012, arXiv:1208.0595
\bibitem[Cohen et al. (2003)]{coh2003} Cohen, M., Wheaton, W.~A., \& Megeath, S.~T.\ 2003, \aj, 126, 1090
\bibitem[Col{\'o}n et al. (2012)]{col2012} Col{\'o}n, K.~D., Ford, E.~B., \& Morehead, R.~C.\ 2012, arXiv:1207.2481
\bibitem[Coughlin et al. (2011)]{cou2011} Coughlin, J.~L., L{\'o}pez-Morales, M., Harrison, T.~E., Ule, N., \& Hoffman, D.~I.\ 2011, \aj, 141, 78
\bibitem[Delgado Mena et al. (2010)]{del2010} Delgado Mena, E., Israelian, G., Gonz{\'a}lez Hern{\'a}ndez, J.~I., et al.\ 2010, \apj, 725, 2349
\bibitem[Dotter et al. (2008)]{dot2008} Dotter, A., Chaboyer, B., Jevremovi{\'c}, D., et al.\ 2008, \apjs, 178, 89 
\bibitem[Eisenstein et al. (2011)]{eis2011} Eisenstein, D.~J., Weinberg, D.~H., Agol, E., et al.\ 2011, \aj, 142, 72
\bibitem[Ford et al. (2012)]{ford2012} Ford, E.~B., Ragozzine, D., Rowe, J.~F., et al.\ 2012, \apj, 756, 185 
\bibitem[Fortney (2012)]{for2012} Fortney, J.~J.\ 2012, \apjl, 747, L27
\bibitem[Fukugita et al. (1996)]{fuk1996} Fukugita, M., Ichikawa, T., Gunn, J.~E., et al.\ 1996, \aj, 111, 1748
\bibitem[Garc{\'{\i}}a P{\'e}rez et al. (2015)]{gar2015} Garc{\'{\i}}a P{\'e}rez, A.~E. et al.\ 2015, in prep
\bibitem[Ghezzi et al. (2010)]{ghe2010} Ghezzi, L., Cunha, K., Smith, V.~V., et al.\ 2010, \apj, 720, 1290
\bibitem[Gilliland et al. (2011)]{gil2011} Gilliland, R.~L., Chaplin, W.~J., Dunham, E.~W., et al.\ 2011, \apjs, 197, 6
\bibitem[Girardi et al. (2012)]{gir2012} Girardi, L., Barbieri, M., Groenewegen, M.~A.~T., et al.\ 2012, Red Giants as Probes of the Structure and Evolution of the Milky Way, 165
\bibitem[Gonz{\'a}lez Hern{\'a}ndez et al. (2010)]{gon2010} Gonz{\'a}lez Hern{\'a}ndez, J.~I., Israelian, G., Santos, N.~C., et al.\ 2010, \apj, 720, 1592
\bibitem[Gunn et al. (2006)]{gun2006} Gunn, J.~E., et al.\ 2006, \aj, 131, 2332
\bibitem[Hirano et al. (2012)]{hir2012} Hirano, T., Sanchis-Ojeda, R., Takeda, Y., et al.\ 2012, \apj, 756, 66
\bibitem[Holman et al. (2010)]{hol2010} Holman, M.~J., Fabrycky, D.~C., Ragozzine, D., et al.\ 2010, Science, 330, 51
\bibitem[Howard et al. (2012)]{how2012} Howard, A.~W., Marcy, G.~W., Bryson, S.~T., et al.\ 2012, \apjs, 201, 15
\bibitem[Howell et al. (2011)]{how2011} Howell, S.~B., Everett, M.~E., Sherry, W., Horch, E., \& Ciardi, D.~R.\ 2011, \aj, 142, 19
\bibitem[Huber et al. (2013)]{hub2013} Huber, D., Carter, J.~A., Barbieri, M., et al.\ 2013, Science, 342, 331
\bibitem[Huber et al. (2014)]{hub2014} Huber, D., Silva Aguirre, V., Matthews, J.~M., et al.\ 2014, \apjs, 211, 2
\bibitem[Kraus \& Hillenbrand (2007)]{kra2007} Kraus, A.~L., \& Hillenbrand, L.~A.\ 2007, \aj, 134, 2340 
\bibitem[Johnson et al. (2010)]{joh2010} Johnson, J.~A., Aller, K.~M., Howard, A.~W., \& Crepp, J.~R.\ 2010, \pasp, 122, 905
\bibitem[Johnson et al. (2011)]{joh2011} Johnson, J.~A., Clanton, C., Howard, A.~W., et al.\ 2011, \apjs, 197, 26
\bibitem[Kilic et al. (2012)]{kil2012} Kilic, M., Brown, W.~R., Allende Prieto, C., et al.\ 2012, \apj, 751, 141
\bibitem[Koch et al. (2010)]{koc2010} Koch, D.~G., Borucki, W.~J., Basri, G., et al.\ 2010, \apjl, 713, L79
\bibitem[Lagrange et al. (2009)]{lag2009} Lagrange, A.-M., Desort, M., Galland, F., Udry, S., \& Mayor, M.\ 2009, \aap, 495, 335
\bibitem[Mahadevan et al. (2015)]{mah2015} Mahadevan, S., Deshpande, R., Bender, C., et al.\ 2015, in prep
\bibitem[Majewski et al. (2010)]{maj2010} Majewski, S.~R., Wilson, J.~C., Hearty, F., Schiavon, R.~R., \& Skrutskie, M.~F.\ 2010, IAU Symposium, 265, 480
\bibitem[Majewski et al. (2015)]{maj2015} Majewski, S.~R., et al.\ 2015, in prep
\bibitem[Maldonado et al. (2013)]{mal2013} Maldonado, J., Villaver, E., \& Eiroa, C.\ 2013, \aap, 554, AA84 
\bibitem[Marcy \& Butler (2000)]{mar2000} Marcy, G.~W. \& Butler, R.~P.\ 2000, \pasp, 112, 137
\bibitem[Mayor et al. (2003)]{may2003} Mayor, M., Pepe, F., Queloz, D., et al.\ 2003, The Messenger, 114, 20
\bibitem[Mayor et al. (2011)]{may2011} Mayor, M., Marmier, M., Lovis, C., et al.\ 2011, arXiv:1109.2497
\bibitem[Mel{\'e}ndez et al. (2009)]{mel2009} Mel{\'e}ndez, J., Asplund, M., Gustafsson, B., \& Yong, D.\ 2009, \apjl, 704, L66
\bibitem[M{\'e}sz{\'a}ros et al. (2012)]{mes2012} M{\'e}sz{\'a}ros, S., Allende Prieto, C., Edvardsson, B., et al.\ 2012, arXiv:1208.1916 
\bibitem[Moorhead et al. (2011)]{moo2011} Moorhead, A.~V., Ford, E.~B., Morehead, R.~C., et al.\ 2011, \apjs, 197, 1
\bibitem[Morton (2012)]{mor2012} Morton, T.~D.\ 2012, arXiv:1206.1568
\bibitem[Morton \& Johnson (2011)]{mor2011} Morton, T.~D., \& Johnson, J.~A.\ 2011, \apj, 738, 170
\bibitem[Nidever et al. (2015)]{nid2015} Nidever, D.~L., Holtzman, J.~A., Allende Prieto, C., et al.\ 2015, arXiv:1501.03742
\bibitem[Nissen (2013)]{nis2013} Nissen, P.~E.\ 2013, \aap, 552, AA73
\bibitem[Ofir et al. (2012)]{ofi2012} Ofir, A., Gandolfi, D., Buchhave, L., et al.\ 2012, \mnras, 423, L1
\bibitem[Perruchot et al. (2008)]{per2008} Perruchot, S., Kohler, D., Bouchy, F., et al.\ 2008, \procspie, 7014, 17
\bibitem[Petigura \& Marcy (2011)]{pet2011} Petigura, E.~A., \& Marcy, G.~W.\ 2011, \apj, 735, 41 
\bibitem[Pr{\v s}a et al. (2011)]{prs2011} Pr{\v s}a, A., Batalha, N., Slawson, R.~W., et al.\ 2011, \aj, 141, 83
\bibitem[Rhodes \& Budding (2014)]{rho2014} Rhodes, M.D. \& Budding, E.\ 2014, \emph{Ap\&SS}
\bibitem[Sahlmann et al. (2011)]{sah2011} Sahlmann, J., et al.\ 2011, \aap, 525, 95
\bibitem[Santerne et al. (2012)]{san2012} Santerne, A., D{\'{\i}}az, R.~F., Moutou, C., et al.\ 2012, \aap, 545, AA76
\bibitem[Santerne et al. (2013)]{san2013} Santerne, A., Fressin, F., D{\'{\i}}az, R.~F., et al.\ 2013, \aap, 557, AA139
\bibitem[Schuler et al. (2011)]{sch2011} Schuler, S.~C., Flateau, D., Cunha, K., et al.\ 2011, \apj, 732, 55
\bibitem[Slawson et al. (2011)]{sla2011} Slawson, R.~W., Pr{\v s}a, A., Welsh, W.~F., et al.\ 2011, \aj, 142, 160
\bibitem[Sousa et al. (2008)]{sou2008} Sousa, S.~G., Santos, N.~C., Mayor, M., et al.\ 2008, \aap, 487, 373
\bibitem[Sousa et al. (2011)]{sou2011} Sousa, S.~G., Santos, N.~C., Israelian, G., Mayor, M., \& Udry, S.\ 2011, \aap, 533, AA141 
\bibitem[Steffen et al. (2010)]{ste2010} Steffen, J.~H., Batalha, N.~M., Borucki, W.~J., et al.\ 2010, \apj, 725, 1226
\bibitem[Steffen et al. (2013)]{ste2013} Steffen, J.~H., Fabrycky, D.~C., Agol, E., et al.\ 2013, \mnras, 428, 1077
\bibitem[Szentgyorgyi \& Fur{\'e}sz(2007)]{sze2007} Szentgyorgyi, A.~H., \& Fur{\'e}sz, G.\ 2007, Revista Mexicana de Astronomia y Astrofisica Conference Series, 28, 129
\bibitem[Teske et al. (2014)]{tes2014} Teske, J.~K., Cunha, K., Smith, V.~V., Schuler, S.~C., \& Griffith, C.~A.\ 2014, \apj, 788, 39
\bibitem[Teske et al. (2015)]{tes2015} Teske, J.~K., Ghezzi, L., Cunha, K., et al.\ 2015, arXiv:1501.02167
\bibitem[Torres et al. (2011)]{tor2011} Torres, G., Fressin, F., Batalha, N.~M., et al.\ 2011, \apj, 727, 24
\bibitem[Tull (1998)]{tul1998} Tull, R.~G.\ 1998, \procspie, 3355, 387
\bibitem[Twicken et al. (2010)]{twi2010} Twicken, J.~D., Clarke, B.~D., Bryson, S.~T., et al.\ 2010, \procspie, 7740, 69
\bibitem[Vogt et al. (1994)]{vog1994} Vogt, S.~S., Allen, S.~L., Bigelow, B.~C., et al.\ 1994, \procspie, 2198, 362
\bibitem[Wang et al. (2014)]{wan2014} Wang, J., Fischer, D.~A., Xie, J.-W., \& Ciardi, D.~R.\ 2014, \apj, 791, 111
\bibitem[Wilson et al. (2010)]{wil2010} Wilson, J.~C., Hearty, F., Skrutskie, M.~F., et al.\ 2010, \procspie, 7735, 77351C
\bibitem[Wilson et al. (2012)]{wil2012} Wilson, J.~C., Hearty, F., Skrutskie, M.~F., et al.\ 2012, \procspie, 8446
\bibitem[Wizinowich et al. (2000)]{wiz2000} Wizinowich, P.~L., Acton, D.~S., Lai, O., et al.\ 2000, \procspie, 4007, 2 
\bibitem[Wright \& Howard (2009)]{wri2009} Wright, J.~T., \& Howard, A.~W.\ 2009, \apjs, 182, 205
\bibitem[Wright et al. (2013)]{wri2013} Wright, J.~T., Roy, A., Mahadevan, S., et al.\ 2013, \apj, 770, 119
\bibitem[Youdin (2011)]{you2011} Youdin, A.~N.\ 2011, \apj, 742, 38
\bibitem[Zamora et al. (2015)]{zam2015} Zamora, O., Garc{\'{\i}}a-Hern{\'{a}}ndez, D. A., Allende Prieto, C., et al.\ 2015, in prep
\bibitem[Zasowski et al. (2013)]{zas2013} Zasowski, G., Johnson, J.~A., Frinchaboy, P.~M., et al.\ 2013, \aj, 146, 81 
\bibitem[Zucker et al. (1995)]{zuc1995} Zucker, S., Torres, G., \& Mazeh, T.\ 1995, \apj, 452, 863
\bibitem[Zucker (2003)]{zuc2003} Zucker, S.\ 2003, \mnras, 342, 1291
\end{thebibliography}
\end{document}